\documentstyle[aps,prb,epsfig,psfig,floats]{revtex}

\begin{document}

%\baselineskip 0.2truecm
\newcommand{\beq}{\begin{equation}}
\newcommand{\eeq}{\end{equation}}
\newcommand{\bea}{\begin{eqnarray}}
\newcommand{\eea}{\end{eqnarray}}
\newcommand{\dfrac}{\displaystyle\frac}                                    
\newcommand{\disp}{\displaystyle}                                    
\newcommand{\mbf}{\mathbf}
\newcommand{\dint}{\int}
\renewcommand{\u}{\underline}                                        
\renewcommand{\o}{\overline}                                        
\newcommand{\Ptwo}{{\mathcal P}_2}
\newcommand{\bi}{\begin{itemize}}
\newcommand{\ei}{\end{itemize}}
\newcommand{\thint}{\widetilde{d\theta}}
\newcommand{\thprint}{\widetilde{d\theta'}}
\newcommand{\angint}{\int \thint\ }
\newcommand{\totint}{\int dl\ \thint\ }
%{\dint \dfrac{1}{2}\ d\cos\theta\ }
\newcommand{\fexc}{\tilde{f}}
\newcommand{\fmom}{f_{\rm{mom}}}
\newcommand{\fid}{f_{\rm{id}}}
\newcommand{\rhoa}{\rho^{(a)}}
\newcommand{\rhob}{\rho^{(b)}}
\newcommand{\va}{v^{(a)}}
\newcommand{\vb}{v^{(b)}}
\newcommand{\parent}{\rho^{(0)}}
\newcommand{\normparent}{P^{(0)}}
\newcommand{\ssch}{\sigma_{\rm S}}
\newcommand{\I}{^{\rm I}}
\newcommand{\N}{^{\rm N}}

\newcommand{\bfig}{\begin{figure}[htb]\begin{center}}
\newcommand{\efig}{\end{center}\end{figure}}
\newcommand{\eqref}[1]{(\ref{#1})}
\newcommand{\eqq}[2]{(\ref{#1},\ref{#2})}
\newcommand{\eqqq}[2]{(\ref{#1}-\ref{#2})}

\newtheorem{thm}{Theorem}     [section]                                    
\newtheorem{definition}{Def.}   [section]                                    
\newtheorem{obs}{Remark}  [section]

\twocolumn[\hsize\textwidth\columnwidth\hsize\csname @twocolumnfalse\endcsname

\title{Simplified Onsager theory for isotropic-nematic phase
equilibria of length polydisperse hard rods}

\author{Alessandro Speranza, Peter Sollich}

\address{Department of Mathematics, King's College London,
Strand, London WC2R 2LS, U.K.
Email: alessandro.speranza@kcl.ac.uk}

\maketitle
\begin{abstract}
Polydispersity is believed to have important effects on the formation
of liquid crystal phases in suspensions of rod-like particles. To
understand such effects, we analyse the phase behaviour of thin hard
rods with length polydispersity. Our treatment is based on a
simplified Onsager theory, obtained by truncating the series expansion
of the angular dependence of the excluded volume.  We
describe the model and give the full phase equilibrium equations;
these are then solved numerically using the {\em moment free energy
method}
which reduces the problem from one with an infinite number of
conserved densities to one with a finite number of effective densities
that are moments of the full density distribution. The method yields
exactly the onset of nematic ordering. Beyond this, results are
approximate but we show that they can be made essentially arbitrarily
precise by adding adaptively chosen extra moments, while still
avoiding the numerical complications of a direct solution of the full
phase equilibrium conditions.

We investigate in detail the phase behaviour of systems with three
different length distributions: a (unimodal) Schulz distribution, a
bidisperse distribution and a bimodal mixture of two Schulz
distributions which interpolates between these two cases.
A three-phase isotropic-nematic-nematic coexistence region is shown to
exist for the bimodal and bidisperse length distributions if the ratio
of long and short rod lengths is sufficiently large, but not for the
unimodal one. We systematically explore the topology of the phase
diagram as a function of the width of the length distribution and of
the rod length ratio in the bidisperse and bimodal cases.
\end{abstract}

%\vfill\eject
\vspace*{0.5cm}
]
%\twocolumn

\section{Introduction}
Suspensions of rod-like particles can undergo an orientational
disorder-order phase transition from a disordered isotropic
(I) phase, in which rods point with equal probability in every
direction, to an orientationally ordered nematic (N) phase in which
particle orientations cluster around a preferred direction. This
behaviour has been observed experimentally in both
chemical~\cite{CopFre37,Langmuir38,NakHayOhm68,Zocher25} and
biological~\cite{Langmuir38,BawPir37,BawPir38,BawPirBerFan36,BerFan37} systems.

Two main theoretical approaches have been used to explain the I-N
phase transition.  The Maier-Saupe theory~\cite{MaiSau58} neglects
density variations and is therefore appropriate for so-called
thermotropic materials, where the transition is driven by variations
in temperature. It focuses on attractive, long-range interactions
between particles; in the original theory these attractions were
thought of as arising from Van der Waals forces but in fact the
Maier-Saupe approach can be considered a lowest order approximation
for general orientation-dependent attractive
interactions~\cite{Frenkel_review}. 
The Onsager theory~\cite{Onsager49}, on
the other hand, takes into account only hard core repulsions between
the particles on contact; temperature then becomes unimportant and so
the theory is appropriate for lyotropic materials, where the phase
ordering is driven by changes in density.
The I-N phase transition arises from the
competition between the orientational entropy (corresponding to the
tendency of rods to stay orientationally disordered) and the packing
entropy (which is higher for aligned rods due to the excluded volume
interaction). 
The key simplification is that, in the
``Onsager limit'' of infinitely thin rods, the virial expansion for
the free energy truncates after the second-order contribution.  The
state of the system is described by the density of rods and the
distribution $P(\theta)$ of the angles $\theta$ which the rods make
with the nematic axis, i.e., the preferred orientation in the nematic phase.
A formal minimization of the free energy with respect to this function
leads to a self-consistency equation for $P(\theta)$. Solving this
gives, in principle, the free energy as a function of density, from
which the I-N phase coexistence region can then be obtained by a
standard double-tangent construction. Onsager gave an approximate
solution for the I-N phase
transition using a variational ansatz for $P(\theta)$. With two
different approximation 
techniques Lekkerkerker et al.~\cite{LekCouVanDeb84} and
Lasher~\cite{Lasher70} obtained more accurate results; the numerically
exact solution has also been 
obtained~\cite{LekCouVanDeb84,KayRav77}.
Isihara~\cite{Isihara51}, basing his
calculations essentially on the Onsager theory, had shown earlier that the
order-disorder phase transition 
will occur at similar densities for particles of different shapes, as
long as they are strongly asymmetric (e.g.\ plates, cylinders,
spheroids).

We mention briefly an alternative to Onsager theory that was
developed by 
Flory~\cite{Flory56_2}, Di Marzio~\cite{DiMarzio61} and Alben~\cite{Alben71}, based on a lattice
model. The interaction is again of excluded volume
type, but particles are now embedded in a three-dimensional cubic
lattice whose sites are occupied either by segments of the rod-like
particle or by solvent. Although qualitetively this model gives
the same behaviour as the Onsager theory, the I-N phase
transition is predicted at a density about twice that found by
Onsager~\cite{Flory56_2,BirKolPry88}. With a similar lattice model
Flory~\cite{Flory56} predicted an analogous order-disorder phase transition
for semiflexible
macromolecules. On the experimental side, Robinson~\cite{Robinson56},
Hermans~\cite{Hermans62} and Nakajima et al.~\cite{NakHayOhm68} found
results in good agreement with Flory's theory for hard rod-like molecules.

Returning now to Onsager theory, a direct comparison between the
theoretical predictions and
experimental data is difficult due to the complexity of experimental
systems. For example, there are corrections due to non-hard
interactions in real systems~\cite{BuiPatPhiLek93,BuiVelPatLek92}
which affect the densities of coexisting isotropic and nematic phases and 
other characteristics of the phase
transition~\cite{KhoSem81,KhoSem81_2}. These effects can be partially
accounted for by adding an
attractive potential to the excluded volume
interaction~\cite{BarGel77,Cotter77}, which leads to a temperature
dependence of e.g.\ the orientational order parameter.

More importantly for our
purposes, polydispersity -- the presence of an (effectively)
continuous spread of rod lengths and/or diameters -- has been shown to
have important effects on the phase
diagram~\cite{BuiLek93,VanVanLek96}. Onsager~\cite{Onsager49} already
gave the extension of his approach to the case of mixtures of  
rods with different lengths or different diameters, but the obvious
complication of having many (and in the fully polydisperse limit, 
infinitely many) distinguishable particle species prevented a
direct numerical solution, or the systematic use of trial functions as
in the monodisperse case. We focus in this paper on {\em length}
polydispersity alone, assuming that rod {\em diameters} are still
monodisperse. Such length polydispersity can lead to a strong
broadening of the coexistence region, which has been observed
experimentally~\cite{BuiLek93,VanVanLek96} and also predicted
theoretically, at least within an expansion for narrow length
distributions~\cite{chen94,Sluckin89}. Coexistence of several nematic
phases (N-N), occasionally with a third isotropic phase (I-N-N), has
also been seen experimentally~\cite{BuiLek93} and been predicted by
theory for bi- and tridisperse systems (comprising rods of two and
three different lengths,
respectively)~\cite{LekCouVanDeb84,OdiLek85,VanMul96,VroLek93}.

In a series of papers by Flory
et al.~\cite{AbeFlo78,FloAbe78,FloFro78,FroFlo78}, the length-polydisperse
case was solved within the lattice model for a bidisperse system as
well as for cases with many different rod lengths (which in the
lattice model are always multiples of a given unit length). The
results obtained are qualitatively similar to the ones 
predicted by the Onsager theory~\cite{BirKolPry88}, with a three-phase
I-N-N region and an N-N region at high density for the bidisperse
system, and with strong fractionation and broadening of the
coexistence region for length distributions of e.g.\
Poisson~\cite{FroFlo78} and exponential~\cite{FloFro78} type.
Moscicki et
al.\ applied the same theory to a system with a Gaussian length
distribution~\cite{MosWil82}, observing strong fractionation and a
pronounced dependence of the size of the coexistence region on the width of the
distribution.

The theoretical studies referred to above which concern themselves
directly with the off-lattice situation are restricted to either
mixtures of rods of two or three different lengths or to continuous
length distributions of small width.
The question of the effects of full length polydispersity within
the Onsager theory of thin hard rods therefore remains open, and it is
this problem that we address in the present paper. It is clear that
the introduction of full polydispersity significantly complicates the
analysis of phase behaviour. Each rod length now has its own conserved
density associated with it, and a naive application of the Gibbs phase
rule suggests that one could have arbitrarily many coexisting
phases. This problem is compounded by the presence of the varying rod
orientations, so that the full density distribution over lengths and
orientations has both conserved and non-conserved parts. A direct
attack on the Onsager model with full length polydispersity is
therefore extremely difficult.

As a first step towards understanding the phase behaviour of
polydisperse hard rods, we have previously analysed the Zwanzig model,
a discretization of the Onsager theory which allows rods to point only
in one of three orthogonal
directions~\cite{Zwanzig63,ClaMcl92,ClaCueSeaSolSpe00}. We found that this
exhibited some of the features that are typical of polydisperse
systems (broadening of the coexistence region, fractionation of rod
lengths across different phases). It was not, however, able to predict
the observed N-N and I-N-N phase separations. This could be traced
back to the restriction on rod orientations: a separation of a single
nematic phase into two nematics containing predominantly shorter and
longer rods involves a loss of entropy of mixing and a gain in
orientational entropy; with only three different 
rod orientations available, the
possible gain in orientational entropy is too limited for such a phase
separation to be favourable.

In the present work we analyse a model which can take the complexities
of full length polydispersity into account and still predict the more
complex phase coexistences involving several nematic phases that were
missing in the Zwanzig model. Our ``$\Ptwo$ Onsager model'' is an
approximation of the full Onsager theory obtained by truncating the
expansion of the angular dependence of the excluded volume of two rods
after the lowest nontrivial term, which involves the Legendre
polynomial $\Ptwo$. The model is thus no longer exact, even in the
limit of thin rods, but nevertheless useful as a solvable
approximation to the full Onsager theory which can, for example,
provide some information about the possible topologies of the phase
diagram. The errors introduced by the truncation can be directly
assessed for the bidisperse case, where accurate numerical solutions
of the full Onsager theory exist~\cite{LekCouVanDeb84,VroLek93}, and
we find below that there the $\Ptwo$ Onsager model reproduces most of the
trends in the evolution of the phase diagram with the rod length
distribution. The truncation errors could in
principle be reduced by extending the angular expansion to
increasingly high orders, giving a series of models whose behaviour
will approach that of the full Onsager theory in the limit; the
$\Ptwo$ Onsager model is the first member of this series. Its free
energy also bears a formal similarity to that of the polydisperse
Maier-Saupe model, and by imposing the constraint of constant density
could yield qualitative information about the phase behaviour of
polydisperse thermotropics. A final benefit is that the $\Ptwo$
Onsager model is, in contrast to the full Onsager theory, {\em
truncatable}: its excess free energy only depends on a finite number
(two) of moments of the density distribution. Phase equilibria can
therefore be found efficiently using the moment free energy
method~\cite{SolCat98,SolWarCat01,Warren98} (or
moment method for short), with
its extension to systems with non-conserved densities introduced in
the context of the polydisperse Zwanzig
model~\cite{ClaCueSeaSolSpe00}.

This paper is structured as follows. In Sec.~\ref{sec:model} we
describe our truncated ``$\Ptwo$ Onsager model'' and its link to the
full Onsager theory, and give the exact phase coexistence equations
for the truncated model. In Sec.~\ref{sec:moment}, we briefly review
the moment method and then apply it to the $\Ptwo$ Onsager model.  The
numerical method used for calculating phase equilibria is described in
some detail, including our procedure for ensuring that the moment method
produces numerically accurate predictions for phase equilibria even
beyond the point where it is exact by construction (which turns
out to be the 
onset of nematic phase ordering). Our main results are detailed in
Sec.~\ref{sec:results}, where we give phase diagrams for three
different types of rod length distributions: Schulz distributions,
which are unimodal, bidisperse distributions containing only two
different rod lengths, and finally mixtures of Schulz distributions
that interpolate between these two extremes. One of our guiding
questions will be under which conditions on the rod length
distribution the phase diagram shows the more ``exotic'' features of
nematic re-entrance and of I-N-N and N-N coexistence regions. We will
find that, while these features are absent for the unimodal Schulz
distributions with even the largest widths that we consider, the
bidisperse and bimodal systems give a much richer phase behaviour.  We
summarize in Sec.~\ref{sec:conclusion} and discuss directions for
future work. App.~\ref{sec:SNO_approx} contains a description of a
simple approximation that we found useful in elucidating certain
trends in the evolution of the phase diagram, particularly in the
limit of large length ratios between long and short rods for the
bimodal and bidisperse distributions.

\section{Model details}
\label{sec:model}

We consider a system of hard cylinders of length $L$ and diameter $D$,
capped with hemispheres at both ends. We assume that these rod-like
particles are polydisperse in their length $L$, but all have identical
diameter $D$. To be able to take the Onsager limit of long thin rods,
we introduce a reference length $L_0$ and the normalized rod lengths
$l=L/L_0$; the Onsager limit is then reached for $D/L_0\rightarrow 0$
at fixed values of $l$.

The state of a single phase of a system is described by the density
distribution $\rho(l,\Omega)$, defined so that
$\rho(l,\Omega)\,dl\,d\Omega/4\pi$ gives the density of rods with
(normalized) lengths in an interval $dl$ around $l$, and orientations
$\Omega$ in a solid angle $d\Omega$. The distribution $\rho(l,\Omega)$
is the natural extension for a polydisperse system of the usual
orientational distribution $P(\Omega)$ used for monodisperse systems
of rods~\cite{VroLek92}. The rod orientation $\Omega$ can be
parameterized in terms of the angle $\theta$ with the nematic axis and
an azimuthal angle $\varphi$; due to the cylindrical symmetry of the
nematic phase $\rho(l,\Omega)$ is independent of $\varphi$. Using
$d\Omega = d\cos\theta\ d\varphi$ the
density distribution as a function of rod length, obtained by
integrating over orientations, is therefore 
\begin{equation}\label{eq:normalization_ang}
\rho(l)=\frac{1}{4\pi}\dint d\Omega\ \rho(l,\Omega)=\angint
\rho(l,\theta)
\end{equation}
Here and below we use the shorthand
\begin{equation}
\angint F(\theta)=\dfrac{1}{2}\dint_{-1}^1 d\cos\theta\ F(\theta)
\end{equation}
for the angular integral of an arbitrary function $F(\theta)$. The
orientational distribution of rods can be factored out from
$\rho(l,\theta)$ as~\cite{ClaCueSeaSolSpe00}
\begin{equation}\label{eq:decomposition_rho}
\rho(l,\theta)=
\rho(l)P_l(\theta)
%=\rho_0 P(l) P_l(\theta)
\end{equation}
where $P_l(\theta)$ represents the probability of finding a rod of
given length $l$ in orientation $\Omega=(\theta,\varphi)$ and is
normalized to $1$:
\begin{equation}\label{eq:normalization_ang_P}
\angint P_l(\theta)=1
\end{equation}
In the isotropic phase, one has $P_l(\theta)\equiv 1$ and
$\rho(l,\theta)=\rho(l)$.

\subsection{Polydisperse Onsager theory}

As mentioned above, the present model is a truncation of the full
Onsager theory (see e.g.\ Refs.~\onlinecite{Sluckin89,VroLek92} for
reviews), which we now describe. In the Onsager limit $D/L_0\to 0$,
the second order virial approximation becomes exact and gives for the
excess free energy in units of $k_{\rm B}T$
\begin{equation}
\label{eq:fexc_first}
\tilde{f}= \dint dl\ dl^\prime\ \dfrac{d\Omega}{4\pi}\ \dfrac{d\Omega^\prime}{4\pi}\
B_2(lL_0,\Omega,l'L_0,\Omega^\prime)
\rho(l, \theta)\rho(l^\prime, \theta^\prime)
\end{equation}
Here $B_2$ is the second virial coefficient which for hard rods is
half the excluded volume and given by 
\begin{equation}\label{eq:B_2}
B_2=DL_0^2\, ll'\,|\sin\gamma|
\end{equation}
up to negligible terms of order $D/L_0$, with $\gamma$ the angle
between the two rods.  Using the cylindrical symmetry of the nematic
phase one can perform the integration over $\varphi$ and
$\varphi^\prime$ in Eq.~(\ref{eq:fexc_first}). If we also make the
densities $\fexc$ and $\rho(l,\theta)$ non-dimensional by multiplying
by a unit volume $(\pi/4)DL_0^2$ we obtain
\begin{equation}
\fexc=\frac{4}{\pi}\int dl\ dl^\prime\ \thint\ \thprint\
ll^\prime\,
K(\theta,\theta^\prime)\,\rho(l,\theta)\rho(l^\prime,\theta^\prime)
\end{equation}
where the angular part,
\begin{eqnarray}\label{eq:kernel}
K(\theta,\theta^\prime)&=&
\dint_0^{2\pi}\dfrac{d\varphi}{2\pi}\ \dfrac{d\varphi^\prime}{2\pi}\
|\sin\gamma| \nonumber\\
&=&\sum_{n=0}^{\infty}k_{2n}{\mathcal P}_{2n}(\cos\theta){\mathcal
P}_{2n}(\cos\theta^\prime)
\end{eqnarray}
can be expressed as a bilinear expansion in Legendre
polynomials~\cite{KayRav77}
${\mathcal P}_{2n}(\cos\theta)$ with $k_0=\pi/4,
k_2=-5\pi/32$ etc. and ${\mathcal P}_0(\cos\theta)=1$,
$\Ptwo(\cos\theta)=\frac{1}{2}(3\cos^2\theta-1)$. The excess free energy is therefore an infinite sum of moments of the
density distribution $\rho(l,\theta)$, with weight functions $l{\mathcal P}_{2n}(\cos\theta)$:
\begin{equation}\label{eq:excess_onsager_fin}
\tilde{f}=\frac{4}{\pi}
\sum_{n=0}^\infty k_{2n}\left(\dint dl\ \thint\ l\,
{\mathcal P}_{2n}(\cos\theta)\rho(l,\theta)\right)^2
\end{equation}

\subsection{The $\Ptwo$ Onsager model}

From Eq.~\eqref{eq:excess_onsager_fin} we see that the full Onsager
theory does not give a {\em truncatable} excess free
energy~\cite{SolWarCat01}, i.e.\ one that is just a function of a
finite number of moments $\rho_i$ of the density distribution
$\rho(l,\theta)$. To obtain a more manageable theory, to which the
efficient moment free energy
method~\cite{SolCat98,SolWarCat01,Warren98} can be applied, we now
truncate the excess free energy after the second Legendre
polynomial. We will thus consider the ``$\Ptwo$ Onsager model''
defined by the excess free energy
\begin{equation}\label{eq:excess_Ptwo}
\fexc=\frac{c_1}{2}\rho_1^2-\frac{c_2}{2}\rho_2^2
\end{equation}
where
\begin{equation}\label{eq:rho1}
\rho_1=\totint l\,\rho(l,\theta)
\end{equation}
and
\begin{equation}\label{eq:rho2}
\rho_2=\totint l\Ptwo(\cos\theta)\, \rho(l,\theta)
\end{equation}
are moments of the density distribution with weight functions 
\begin{eqnarray}
w_1(l,\theta)&=&l\label{eq:w1}\\
w_2(l,\theta)&=&l\Ptwo(\cos\theta)\label{eq:w2}
\end{eqnarray}
and the numerical constants are $c_1=(8/\pi)k_0=2$,
$c_2=-(8/\pi)k_2=5/4$. In our units $\rho_1$ is the rescaled volume
fraction $\phi$ of rods, $\rho_1=(L_0/D) \phi$.  We will denote below by
$\rho_0$ the zeroth moment of the density distribution, $\rho_0=\int
dl\, \rho(l)$, with weight function $w_0(l,\theta)=1$; this is the
number density of rods. The ratio $\rho_1/\rho_0=\langle l\rangle$
then gives the rod length averaged over the normalized length
distribution $P(l)=\rho(l)/\rho_0$, and $\rho_2/\rho_1\leq 1$
expresses the degree of nematic ordering.

It is clear that the truncated model defined above will give
approximate results compared to the predictions of the full Onsager
theory. However, previous work~\cite{LekCouVanDeb84} for the
monodisperse case has investigated the convergence of the densities of
coexisting isotropic and nematic phases when the expansion
\eqref{eq:excess_onsager_fin} is truncated at higher and higher
orders. Already for the truncation after $\Ptwo$ the
results were qualitatively correct, and this encourages us to study
polydispersity effects within the truncated model defined above.

A different rationale for studying the $\Ptwo$ Onsager model could
come from the fact that its excess free energy~\eqref{eq:excess_Ptwo}
is very similar to that of the polydisperse Maier-Saupe model for
thermotropics (see e.g.\ Refs.~\onlinecite{Sluckin89}). There the interaction is 
not of the hard core type, but if one assumes that the interactions of
rods of lengths $l$ and $l'$ again scales as $ll'$ the final
expression for the excess free energy is identical to
Eq.~\eqref{eq:excess_Ptwo}, with $c_1=0$ and $c_2=c_2(T)$ a function
of temperature. The actual predictions of the two theories would
nevertheless not necessarily be close, since within Maier-Saupe theory
the overall rod density $\rho_0$ is assumed to be the same in all
phases; this implies, for example, that a finite region of I-N
coexistence cannot appear for monodisperse rods (see
e.g.\ Ref.~\onlinecite{Sluckin89}). We therefore leave a study of polydisperse
Maier-Saupe theory for future work.

To complete the specification of our model, we need to add to the free
energy the ideal part, i.e.\ the free energy of an ideal mixture of
polydisperse particles (see
e.g.\ Refs.~\onlinecite{Sluckin89,SolCat98,ClaCueSeaSolSpe00})
\begin{equation}\label{eq:fid}
\fid=\angint dl\ \rho(l, \theta)\left[\ln\rho(l, \theta)-1\right]
\end{equation}
Using the decomposition~\eqref{eq:decomposition_rho} and the
relation~\eqref{eq:normalization_ang}, the total free energy (density)
of our model is therefore:
\begin{equation}\label{eq:free_en}
f=\dint dl\ \rho(l)\left[\ln\rho(l)-1\right]+\totint \rho(l)
P_l(\theta)\ln P_l(\theta)+\fexc
\end{equation} 
with $\fexc$ given by~\eqref{eq:excess_Ptwo}.
For a given density distribution
$\rho(l)$, $P_l(\theta)$ is obtained by minimization of
the free energy with respect of $P_l(\theta)$, subject to the
constraint~\eqref{eq:normalization_ang_P}. Introducing corresponding Lagrange
multipliers $\kappa(l)$ one has the condition
\begin{eqnarray}
0 &=& \frac{\delta}{\delta P_l(\theta)}\left(f+\totint \kappa(l)
P_l(\theta)\right)\\
&=&
\rho(l)\left[\ln P_l(\theta)+1\right]+l\rho(l)
\left[c_1\rho_1-c_2\rho_2\Ptwo\right]+\kappa(l)
\end{eqnarray}
Solving for $P_l(\theta)$ gives (here and in the following
$\Ptwo\equiv \Ptwo(\cos\theta)$)
\begin{equation}\label{eq:P_l}
P_l(\theta)=\frac{\exp(lc_2\rho_2\Ptwo)}{\angint
\exp(lc_2\rho_2\Ptwo)}
\end{equation}
The orientational distributions are therefore determined only by
$\rho_2$, which from~\eqref{eq:rho2} obeys the self-consistency
equation
\begin{equation}
\label{eq:rho2_selfconsis}
\rho_2=\totint \rho(l)
\frac{l\Ptwo(\cos\theta)\exp(lc_2\rho_2\Ptwo)}{\angint \exp(lc_2\rho_2\Ptwo)}
\end{equation}
If there are several solutions $\rho_2$ for a given $\rho(l)$, the
one with the smaller free energy~\eqref{eq:free_en} is the physical one.

\subsection{Phase coexistence equations}

At this point we need the expressions for the chemical
potential $\mu(l)$ and the (osmotic) pressure $\Pi$ in order to derive
the phase equilibrium conditions. The chemical potential is obtained
by functional 
differentiation of the free energy~\eqref{eq:free_en} with respect to
$\rho(l)$; the variations of the $P_l(\theta)$ with $\rho(l)$ need not
be considered since $P_l(\theta)$ is chosen to minimize $f$. After
a little simplification one thus finds
\begin{equation}\label{eq:mu}
\mu(l)=\ln\rho(l)-\ln\angint\exp(-lc_1\rho_1+lc_2\rho_2\Ptwo)
\end{equation}
The pressure follows from the Gibbs-Duhem relation as
\begin{equation}\label{eq:Pi}
\Pi=-f+\int dl\
\rho(l)\mu(l)=\rho_0+\frac{c_1}{2}\rho_1^2-\frac{c_2}{2}\rho_2^2
\end{equation}
Imposing equality of chemical potentials in a set of $P$ coexisting
phases, labelled by $a = 1\ldots P$, we can express the density
distributions as
\begin{equation}\label{eq:phase_distrib}
\rhoa(l)
=R(l)\angint\exp(-lc_1\rhoa_1+lc_2\rhoa_2\Ptwo)
\end{equation}
where $R(l)$ is a function of $l$ common to all phases. Using
Eq.~\eqref{eq:P_l}, the full density
distributions over rod lengths and orientations follow as
\beq\label{eq:full_distrib}
\rho^{(a)}(l,\theta) = R(l) \exp(-lc_1\rhoa_1+lc_2\rhoa_2\Ptwo)
\eeq
The function $R(l)$ can be found from Eq.~\eqref{eq:phase_distrib} and
the requirement of particle conservation: if $\parent(l)$ is the
overall or ``parent'' density distribution and $v^{(a)}$ is the
fraction of the system volume occupied by phase $a$, then
\begin{equation}\label{eq:lever_rule}
\sum_a\va\rhoa(l)=\parent(l)
\end{equation}
This leads to
\begin{equation}\label{eq:R(l)}
R(l)=\frac{\parent(l)}{\sum_a\va\angint\exp(-lc_1\rhoa_1+lc_2\rhoa_2\Ptwo)}
\end{equation}
so that, from Eqs.~(\ref{eq:P_l},\ref{eq:phase_distrib}), the density
distributions in the coexisting phases are
\begin{equation}\label{eq:rhoa_l_theta}
\rhoa(l,\theta) = 
\frac{\parent(l) \exp(-lc_1\rhoa_1+lc_2\rhoa_2\Ptwo)}{\sum_b\vb\angint\exp(-lc_1\rhob_1+lc_2\rhob_2\Ptwo)}
\end{equation}
In principle we thus have the conditions for coexistence between $P$
phases: In each phase, $\rho_2$ obeys the self-consistency
condition~\eqref{eq:rho2_selfconsis} and $\rho_1$ a similar equation
obtained by inserting~\eqref{eq:rhoa_l_theta} into~\eqref{eq:rho1};
for the $P$ phase volume fractions $\va$ we have the equality of
pressure in all phases ($P-1$ equations) plus the normalization
$\sum_a \va = 1$. However, finding a starting point from which a
numerical solution of this strongly coupled system of nonlinear
equations converges successfully is very difficult. We use instead the
moment free energy method, which gives exact results for some features
of the phase diagram and otherwise allows us to approach the solution
of the full phase equilibrium conditions with controllable accuracy.

%%%%%%%%%%%%%%%MOMENT FRE EN.%%%%%%%%%%%%%%%%%%%%%%%%

\section{The moment method}
\label{sec:moment}

We only outline the construction of the moment free energy here and
refer to Ref.~\onlinecite{SolWarCat01} for details of its properties,
and Ref.~\onlinecite{ClaCueSeaSolSpe00} for the extension to
non-conserved degrees of freedom such as the rod orientations in our
case. Terms which are linear in the conserved densities $\rho(l)=\int
\thint\ \rho(l,\theta)$ can be added to the free energy~\eqref{eq:free_en}
without affecting the phase equilibria, since they merely add
constants to the chemical potentials $\mu(l)$. We are therefore free
to replace $f$ by
\begin{equation}\label{eq:free_en_2}
f=\angint dl\
\rho(l,\theta)\left[\ln\frac{\rho(l,\theta)}{r(l)}-1\right]+\fexc
\end{equation}
with $r(l)$ an arbitrary function of $l$. Guided by the intuition that
it is the moment densities $\rho_1$ and $\rho_2$ appearing in the
free energy that drive phase separation, we now allow violations of
the lever rule~\eqref{eq:lever_rule} as long as they do not affect
these moments. All other details of the density distribution
$\rho(l,\theta)$ are then found by minimizing the free
energy~\eqref{eq:free_en_2} for the given values of $\rho_1$ and
$\rho_2$ as defined in Eqs.~(\ref{eq:rho1},\ref{eq:rho2}). Inserting a
Lagrange multiplier $\lambda_i$ for each moment $\rho_i$ ($i=1,2$), the
minimum value of the free energy is then
\begin{equation}\label{eq:mom_free_en}
f_{\rm{mom}}=\sum_i\lambda_i\rho_i-\rho_0+\fexc
\end{equation}
where $\rho_0=\int dl\ \rho(l)$ is the number density of rods as
defined previously. The density distribution at which this minimum
free energy is obtained is
\begin{equation}\label{eq:mom_distrib}
\rho(l,\theta)=r(l)\exp\left(\sum_i\lambda_iw_i(l,\theta)\right)
\end{equation}
where the $w_i(l,\theta)$ are the weight functions
(\ref{eq:w1},\ref{eq:w2}) defining the moments; the $\lambda_i$ are
determined implicitly by the requirement that the density
distribution~\eqref{eq:mom_distrib} gives the correct values for the
$\rho_i$. 

Eq.~\eqref{eq:mom_free_en} defines the {\em moment free energy};
rather than being a functional of the density distribution $\rho(l)$,
i.e.\ of an infinite number of conserved densities, it is a
function of the densities $\rho_1$ and $\rho_2$. To find phase
equilibria using the moment free energy, one proceeds as follows.
The ``moment chemical potentials'' are defined as $\mu_i=\partial
f_{\rm{mom}}/\partial\rho_i$ and can be written as~\cite{SolWarCat01}
$\mu_i = \lambda_i + \partial \fexc/\partial\rho_i$, giving in our case
\begin{equation}\label{eq:mu1}
\mu_1=\lambda_1+c_1\rho_1
\end{equation}  
\begin{equation}\label{eq:mu2}
\mu_2=\lambda_2-c_2\rho_2
\end{equation}  
Since $\rho_1$ is conserved while $\rho_2$, which contains the rod
orientations, is not, $\mu_1$ has to be identical in a set of
coexisting phases while $\mu_2$ has to vanish in all the phases. The
pressure calculated from $\fmom$,
\begin{equation}\label{eq:mom_Pi}
\Pi=-\fmom+\rho_1\mu_1+\rho_2\mu_2=\rho_0+
\frac{c_1}{2}\rho_1^2-\frac{c_2}{2}\rho_2^2
\end{equation}
also has to be identical in all phases. Finally, the lever rule has to
be satisfied for the conserved moment $\rho_1$, i.e.\ $\rho_1^{(0)} =
\sum_a v^{(a)} \rho_1^{(a)}$.

It is now easy to see that the moment free energy gives the correct
phase equilibrium conditions, apart from the violations of the lever
rule that it allows. In fact, equality of $\mu_1$ and vanishing
$\mu_2$ imply from Eq.~\eqref{eq:mom_distrib} that the coexisting
phases calculated from the moment free energy can be written as
\beq
\rho^{(a)}(l,\theta)=r(l)\exp\left(\mu_1 l - lc_1 \rho_1^{(a)} -
lc_2\rho_2^{(a)}\Ptwo\right)
\eeq
This is of exactly the form~\eqref{eq:full_distrib} that we derived
earlier from the full phase equilibrium conditions and the
minimization of the free energy w.r.t.\ the rod angle
distributions. The requirement of equal pressures in all phases is
also captured exactly by the moment free energy, since the
expression~\eqref{eq:mom_Pi} actually coincides with the one derived
from the original free energy of the $\Ptwo$ Onsager model,
Eq.~\eqref{eq:Pi}.

We have seen so far that any phase equilibria calculated
from the moment free energy obey the exact conditions of equality of
chemical potentials and pressure. The lever rule will be satisfied
only for $\rho_1$ but not, in general, for the whole of
$\rho(l)$. However, this final requirement will clearly also be
fulfilled {\em if} all phases in an exactly calculated phase split of
the $\Ptwo$ Onsager model are of the form~\eqref{eq:mom_distrib}. We
can guarantee that this is the case at least until the onset of
nematic order (the so-called ``cloud point'' of the isotropic phase),
by choosing the so far unspecified function $r(l)$ to be equal to the
parent density distribution $\parent(l)$. This is the choice we make
from now on. It works because at the isotropic cloud point the nematic
phase still occupies a negligible fraction of the system volume, so
that only the isotropic phase contributes to the denominator of
Eq.~\eqref{eq:rhoa_l_theta}. Explicitly one gets from
Eq.~\eqref{eq:rhoa_l_theta}, with obvious labels
for the phases, $\rho\I(l,\theta)=\parent(l)$ and $\rho\N(l,\theta) =
\parent(l) \exp [lc_1(\rho_1\I-\rho_1\N)+lc_2\rho_2\N\Ptwo]$; as
claimed, these are both of the form~\eqref{eq:mom_distrib} with
$r(l)=\parent(l)$.

Beyond the onset of nematic phase coexistence, phase equilibria
constructed from the moment free energy will in general not solve the
full phase equilibrium conditions, because the lever rule cannot be
satisfied with density distributions from the
family~\eqref{eq:mom_distrib}. Violations of the lever rule can,
however, be minimized by enlarging the family~\eqref{eq:mom_distrib}
to include the actual density distributions occurring in the
coexisting phases.
This can be done by retaining extra moments beyond those appearing in
the excess free energy. Obviously, in the limit of retaining
infinitely many extra moments, the moment free energy would eventually
recover the full free energy of the model, but any computational
efficiency would then be lost. The idea is therefore to add only a few
extra moments to reduce the approximation caused by
using the moment method, keeping their number sufficiently low
to avoid making the computation too slow.
 
In order to enlarge the family~\eqref{eq:mom_distrib} we could in
principle use moments defined by any weight function
$w_i(l,\theta)$. 
Since the corresponding moments $\rho_i$ do not
appear in the excess free energy, their moment chemical potentials are
$\mu_i=\lambda_i + \partial \fexc/\partial \rho_i = \lambda_i$. The
values of the extra Lagrange multipliers must therefore be the same in
all phases, and from Eq.~\eqref{eq:mom_distrib} we can write the
density distributions predicted by the moment method as
\begin{eqnarray}
\label{eq:extra_moms}
\rhoa(l,\theta)&=&\parent(l)\exp\left[\sum_{i=1,2}\lambda^{(a)}_i
w_i(l,\theta)\right]\nonumber\\
&\times&\exp\left[\sum_{i\neq 1,2}\lambda_i w_i(l,\theta)\right]
\end{eqnarray}
Comparing with Eq.~\eqref{eq:rhoa_l_theta}, one sees that the second
exponential, in which the extra weight functions appear, ought to
provide a good approximation to the denominator of
Eq.~\eqref{eq:R(l)}, which is not present
in Eq.~\eqref{eq:mom_distrib}. Since that denominator is a function of $l$
only, the same needs to be true for the extra weight functions. We
will always retain the zeroth moment, i.e.\ the total number density,
with weight function $w_0=1$. This ensures that the ``dilution line''
of density distributions $\rho(l)={\rm const.}\times\parent(l)$ is
included in the family~\eqref{eq:mom_distrib}, a useful feature when
we want to scan across the phase diagram in the direction of
increasing or decreasing density.

In order to choose the other extra weight functions, we apply the
recently developed ``adaptive method''~\cite{ClaCueSeaSolSpe00}. The
intuition behind this is as follows. A phase split calculated by the
moment method for a given point in the phase diagram will give a total
density distribution $\rho_{\rm{tot}}(l)=\sum_a\va\rhoa(l)$ which in
general differs from the parent $\parent(l)$ (though agreeing in the
value of the conserved moment $\rho_1$). A convenient measure of this
violation of the lever rule is the ``log-ratio''
$\ln\rho_{\rm{tot}}(l)/\parent(l)$. Since from
Eq.~\eqref{eq:extra_moms} extra weight functions appear exponentially
in the density distributions, including an extra weight function
$w_3(l)$ similar in shape to the log-ratio should drive $\rho_{\rm
tot}(l)$ closer to $\parent(l)$; it extends the
family~\eqref{eq:mom_distrib} in the ``right direction'' in order to
contain the exact distribution. The method therefore proceeds by
iteratively including extra weight functions that are fitted to the
current log-ratio; a proliferation of extra moments is avoided by
combining the extra weight functions
appropriately~\cite{ClaCueSeaSolSpe00}.
In this way one never requires more than two extra moments in the
calculation, but these are continually adapted to approach the
solution of the full phase equilibrium equations. The quality of the
approximation is measured by the average square
log-ratio~\cite{ClaCueSeaSolSpe00}
\[
\delta=\int dl\
\left(\ln\frac{\rho_{\rm{tot}}(l)}{\parent(l)}\right)^2
\]
and we could easily reach values of $\delta<10^{-6}$, implying that
our results are essentially identical to those that would result from
a direct solution of the full phase equilibrium conditions for the
$\Ptwo$ Onsager model. We verified this explicitly at some points in
the phase diagram: with the solution obtained from our procedure as a
starting point, a Newton-Raphson solver for the full phase equilibrium
conditions converged and terminated at an essentially
indistinguishable solution.

To fit the extra weight functions to the log-ratio in the above
procedure, we initially followed the method of
Ref.~\onlinecite{ClaCueSeaSolSpe00}, representing the weight functions
as linear combinations of some fixed set of basis functions. For some
of the more numerically difficult situations, such as bimodal
distributions with strongly different weights in the two peaks, it
turned out to be more efficient instead to represent the extra weight
functions as spline (piecewise cubic polynomial) fits to the
log-ratio.

\section{Phase diagrams}
\label{sec:results}

In this section we will present our phase diagrams for the length
polydisperse $\Ptwo$ Onsager model. We consider first a Schulz
distribution of lengths, which has frequently been used to model
simple unimodal distributions (see e.g.\
Ref.~\onlinecite{ClaCueSeaSolSpe00}). We will find strong differences
compared to the results for the bi- and tridisperse (full) Onsager
model. By contrasting with the predictions of the $\Ptwo$ Onsager
model for bidisperse systems, we show that these differences are not
due to our truncation of the model but rather due to qualitative
differences in phase behaviour for discrete and continuous length
distributions. Finally, we study a bimodal mixture of two Schulz
distributions. This allows us to analyse in detail how the topology of
the phase diagram changes as we interpolate between the two extremes
of bidisperse and unimodal distributions.

\subsection{Unimodal length distribution}

We study a parent phase with density distribution
$\parent(l)=\parent_0 P^{(0)}(l)$ where the normalized length distribution
is of the Schulz form:
\begin{equation}\label{eq:schulz}
P^{(0)}(l)=\frac{(z+1)^{z+1}}{\Gamma(z+1)}\, l^z\exp[-(z+1)l]
\end{equation}
This gives an average rod length of one for the parent, a convention
we will follow throughout since a different average parent rod length
could always be absorbed into the reference length $L_0$.  The parameter
$z$ controls the width of the distribution, more precisely its
normalized standard deviation $\sigma$, which from now on we will
refer to simply as ``polydispersity'':
\begin{equation}\label{eq:sigma}
\sigma^2=\frac{\langle l^2\rangle - \langle l \rangle^2}{\langle
l\rangle^2}=\frac{1}{z+1}%\frac{1}{\langle l\rangle^2}\int dl\ l^2P^{(0)}(l))-1=
\end{equation} 
The limit $z\rightarrow\infty$ therefore corresponds to the
monodisperse case, where the Schulz distribution degenerates to a
$\delta$-peak. Decreasing $z$ will give increasing polydispersity; we
limit ourselves to $z\geq 0$ and thus $\sigma\leq 1$ since otherwise
the Schulz distribution~\eqref{eq:schulz} has a power-law divergence
for $l\to 0$.
\bfig %\input{phased_new}
\begin{picture}(0,0)%
\epsfig{file=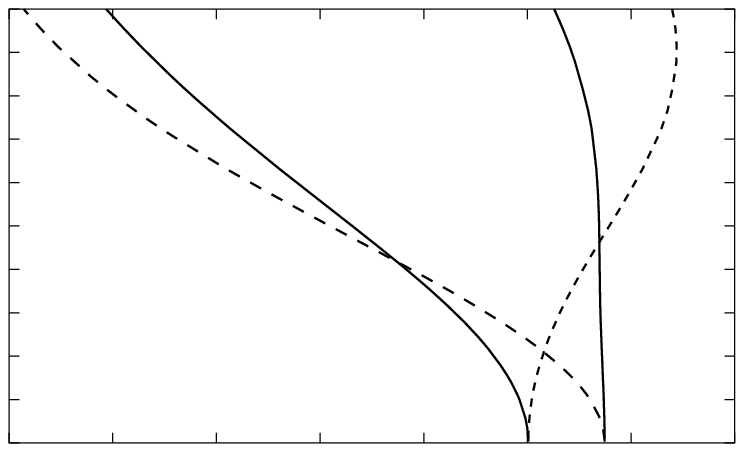}%
\end{picture}%
\setlength{\unitlength}{2565sp}%
\begingroup\makeatletter\ifx\SetFigFont\undefined%
\gdef\SetFigFont#1#2#3#4#5{%
  \reset@font\fontsize{#1}{#2pt}%
  \fontfamily{#3}\fontseries{#4}\fontshape{#5}%
  \selectfont}%
\fi\endgroup%
\begin{picture}(6220,3677)(1,-3211)
\put(3451,-3211){\makebox(0,0)[lb]{\smash{\SetFigFont{8}{9.6}{\familydefault}{\mddefault}{\updefault}$\rho_0$}}}
\put(  1,-511){\makebox(0,0)[lb]{\smash{\SetFigFont{8}{9.6}{\familydefault}{\mddefault}{\updefault}$\sigma$}}}
\put(659,-1914){\makebox(0,0)[rb]{\smash{\SetFigFont{8}{9.6}{\familydefault}{\mddefault}{\updefault}0.3}}}
\put(659,-1593){\makebox(0,0)[rb]{\smash{\SetFigFont{8}{9.6}{\familydefault}{\mddefault}{\updefault}0.4}}}
\put(659,-1273){\makebox(0,0)[rb]{\smash{\SetFigFont{8}{9.6}{\familydefault}{\mddefault}{\updefault}0.5}}}
\put(659,-953){\makebox(0,0)[rb]{\smash{\SetFigFont{8}{9.6}{\familydefault}{\mddefault}{\updefault}0.6}}}
\put(659,-632){\makebox(0,0)[rb]{\smash{\SetFigFont{8}{9.6}{\familydefault}{\mddefault}{\updefault}0.7}}}
\put(659,-312){\makebox(0,0)[rb]{\smash{\SetFigFont{8}{9.6}{\familydefault}{\mddefault}{\updefault}0.8}}}
\put(659,  9){\makebox(0,0)[rb]{\smash{\SetFigFont{8}{9.6}{\familydefault}{\mddefault}{\updefault}0.9}}}
\put(659,329){\makebox(0,0)[rb]{\smash{\SetFigFont{8}{9.6}{\familydefault}{\mddefault}{\updefault}1}}}
\put(659,-2234){\makebox(0,0)[rb]{\smash{\SetFigFont{8}{9.6}{\familydefault}{\mddefault}{\updefault}0.2}}}
\put(1498,-2999){\makebox(0,0)[b]{\smash{\SetFigFont{8}{9.6}{\familydefault}{\mddefault}{\updefault}1.5}}}
\put(659,-2875){\makebox(0,0)[rb]{\smash{\SetFigFont{8}{9.6}{\familydefault}{\mddefault}{\updefault}0}}}
\put(2264,-2999){\makebox(0,0)[b]{\smash{\SetFigFont{8}{9.6}{\familydefault}{\mddefault}{\updefault}2}}}
\put(3029,-2999){\makebox(0,0)[b]{\smash{\SetFigFont{8}{9.6}{\familydefault}{\mddefault}{\updefault}2.5}}}
\put(3795,-2999){\makebox(0,0)[b]{\smash{\SetFigFont{8}{9.6}{\familydefault}{\mddefault}{\updefault}3}}}
\put(4560,-2999){\makebox(0,0)[b]{\smash{\SetFigFont{8}{9.6}{\familydefault}{\mddefault}{\updefault}3.5}}}
\put(5326,-2999){\makebox(0,0)[b]{\smash{\SetFigFont{8}{9.6}{\familydefault}{\mddefault}{\updefault}4}}}
\put(6091,-2999){\makebox(0,0)[b]{\smash{\SetFigFont{8}{9.6}{\familydefault}{\mddefault}{\updefault}4.5}}}
\put(733,-2999){\makebox(0,0)[b]{\smash{\SetFigFont{8}{9.6}{\familydefault}{\mddefault}{\updefault}1}}}
\put(659,-2555){\makebox(0,0)[rb]{\smash{\SetFigFont{8}{9.6}{\familydefault}{\mddefault}{\updefault}0.1}}}
\end{picture}
\caption{Rod number densities of the isotropic and nematic cloud
phases (solid) and of the corresponding shadows (dashed), plotted
against the polydispersity $\sigma$ on the vertical axis. The
isotropic cloud curve has the lower density of the two cloud curves
and joins the isotropic shadow at $\sigma=0$, where the system becomes
monodisperse. The nematic cloud and shadow curves similarly join in
the limit.}
\label{fig:phase_diag_unimodal}
\efig
In Fig.~\ref{fig:phase_diag_unimodal} we show, for each value of
$\sigma$, the isotropic cloud and nematic shadow -- giving the rod
number densities in the coexisting isotropic and nematic phases at the
point where the nematic first appears -- and the nematic cloud and
isotropic shadow, which relate to the point where the isotropic phase
disappears as density is increased.  As anticipated, the coexistence
region, which is delimited by the two cloud curves, broadens strongly
with increasing polydispersity. The other interesting feature, which
was not seen in our earlier study of polydispersity effects in the
Zwanzig model~\cite{ClaCueSeaSolSpe00}, is the crossing between the
cloud curves and the corresponding shadow curves; for values of $\sigma$
above the crossing points the number density of the isotropic phase is
{\em larger} than the one of the coexisting nematic phase.
\bfig
\begin{picture}(0,0)%
\epsfig{file=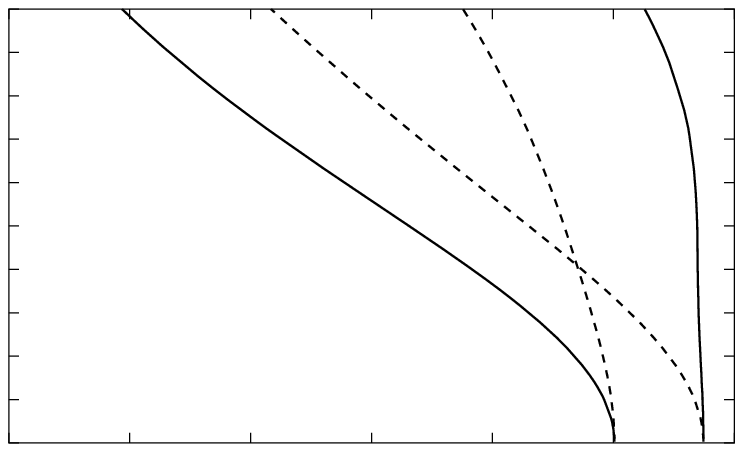}%
\end{picture}%
\setlength{\unitlength}{2565sp}%
\begingroup\makeatletter\ifx\SetFigFont\undefined%
\gdef\SetFigFont#1#2#3#4#5{%
  \reset@font\fontsize{#1}{#2pt}%
  \fontfamily{#3}\fontseries{#4}\fontshape{#5}%
  \selectfont}%
\fi\endgroup%
\begin{picture}(5990,3677)(151,-3211)
\put(3151,-3211){\makebox(0,0)[lb]{\smash{\SetFigFont{8}{9.6}{\familydefault}{\mddefault}{\updefault}$\rho_1$}}}
\put(151,-511){\makebox(0,0)[lb]{\smash{\SetFigFont{8}{9.6}{\familydefault}{\mddefault}{\updefault}$\sigma$}}}
\put(659,-1914){\makebox(0,0)[rb]{\smash{\SetFigFont{8}{9.6}{\familydefault}{\mddefault}{\updefault}0.3}}}
\put(659,-1593){\makebox(0,0)[rb]{\smash{\SetFigFont{8}{9.6}{\familydefault}{\mddefault}{\updefault}0.4}}}
\put(659,-1273){\makebox(0,0)[rb]{\smash{\SetFigFont{8}{9.6}{\familydefault}{\mddefault}{\updefault}0.5}}}
\put(659,-953){\makebox(0,0)[rb]{\smash{\SetFigFont{8}{9.6}{\familydefault}{\mddefault}{\updefault}0.6}}}
\put(659,-632){\makebox(0,0)[rb]{\smash{\SetFigFont{8}{9.6}{\familydefault}{\mddefault}{\updefault}0.7}}}
\put(659,-312){\makebox(0,0)[rb]{\smash{\SetFigFont{8}{9.6}{\familydefault}{\mddefault}{\updefault}0.8}}}
\put(659,  9){\makebox(0,0)[rb]{\smash{\SetFigFont{8}{9.6}{\familydefault}{\mddefault}{\updefault}0.9}}}
\put(659,329){\makebox(0,0)[rb]{\smash{\SetFigFont{8}{9.6}{\familydefault}{\mddefault}{\updefault}1}}}
\put(659,-2234){\makebox(0,0)[rb]{\smash{\SetFigFont{8}{9.6}{\familydefault}{\mddefault}{\updefault}0.2}}}
\put(733,-2999){\makebox(0,0)[b]{\smash{\SetFigFont{8}{9.6}{\familydefault}{\mddefault}{\updefault}1}}}
\put(659,-2875){\makebox(0,0)[rb]{\smash{\SetFigFont{8}{9.6}{\familydefault}{\mddefault}{\updefault}0}}}
\put(1626,-2999){\makebox(0,0)[b]{\smash{\SetFigFont{8}{9.6}{\familydefault}{\mddefault}{\updefault}1.5}}}
\put(2519,-2999){\makebox(0,0)[b]{\smash{\SetFigFont{8}{9.6}{\familydefault}{\mddefault}{\updefault}2}}}
\put(3412,-2999){\makebox(0,0)[b]{\smash{\SetFigFont{8}{9.6}{\familydefault}{\mddefault}{\updefault}2.5}}}
\put(4305,-2999){\makebox(0,0)[b]{\smash{\SetFigFont{8}{9.6}{\familydefault}{\mddefault}{\updefault}3}}}
\put(5198,-2999){\makebox(0,0)[b]{\smash{\SetFigFont{8}{9.6}{\familydefault}{\mddefault}{\updefault}3.5}}}
\put(6091,-2999){\makebox(0,0)[b]{\smash{\SetFigFont{8}{9.6}{\familydefault}{\mddefault}{\updefault}4}}}
\put(659,-2555){\makebox(0,0)[rb]{\smash{\SetFigFont{8}{9.6}{\familydefault}{\mddefault}{\updefault}0.1}}}
\end{picture}
\vspace*{0.2cm}
\caption{The scaled rod volume fractions of the isotropic and nematic
cloud phases (solid) and the relative shadows (dashed), plotted
against the polydispersity $\sigma$.}
\label{fig:vol_frac_unimodal}
\efig
If we switch to a different representation of the phase diagram by plotting the
(scaled) rod volume fractions $\rho_1$ rather than the number
densities $\rho_0$ in the various phases
(Fig.~\ref{fig:vol_frac_unimodal}) the crossing between cloud and
shadow curves disappears. So the rod volume fraction
in the nematic is always larger than in the isotropic phase, even though
the isotropic can have the larger number density. Since
$\rho_1/\rho_0$ is the average rod length in each phase, this is
clear evidence of a strong fractionation effect, with the longer rods being
found predominantly in the nematic phase. We show
this explicitly in Fig.~\ref{fig:fractionation_unimodal}. Bearing in mind
that the cloud phases have the same length distributions as the parent
and therefore an average rod length of unity, one sees that indeed
the average rod length in the nematic shadow is much larger than 
in its corresponding isotropic cloud, and that of the nematic cloud is
much larger than that of the corresponding isotropic shadow.
\bfig
\begin{picture}(0,0)%
\epsfig{file=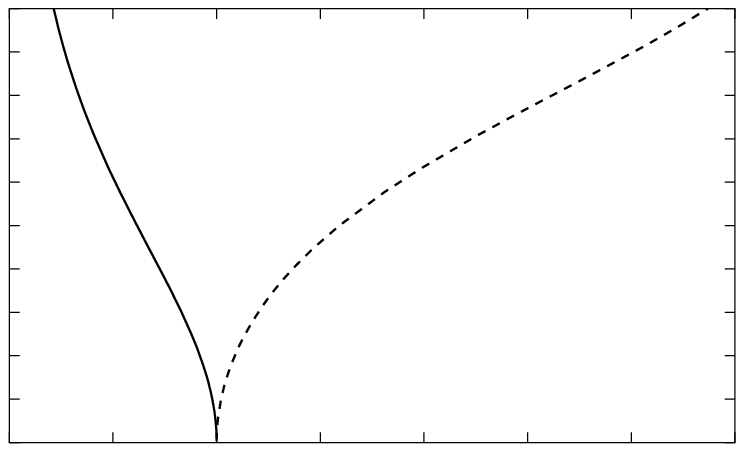}%
\end{picture}%
\setlength{\unitlength}{2565sp}%
\begingroup\makeatletter\ifx\SetFigFont\undefined%
\gdef\SetFigFont#1#2#3#4#5{%
  \reset@font\fontsize{#1}{#2pt}%
  \fontfamily{#3}\fontseries{#4}\fontshape{#5}%
  \selectfont}%
\fi\endgroup%
\begin{picture}(5915,3731)(226,-3265)
\put(226,-511){\makebox(0,0)[lb]{\smash{\SetFigFont{8}{9.6}{\familydefault}{\mddefault}{\updefault}$\sigma$}}}
\put(3376,-3211){\makebox(0,0)[lb]{\smash{\SetFigFont{8}{9.6}{\familydefault}{\mddefault}{\updefault}$\langle l\rangle$}}}
\put(659,-1914){\makebox(0,0)[rb]{\smash{\SetFigFont{8}{9.6}{\familydefault}{\mddefault}{\updefault}0.3}}}
\put(659,-1593){\makebox(0,0)[rb]{\smash{\SetFigFont{8}{9.6}{\familydefault}{\mddefault}{\updefault}0.4}}}
\put(659,-1273){\makebox(0,0)[rb]{\smash{\SetFigFont{8}{9.6}{\familydefault}{\mddefault}{\updefault}0.5}}}
\put(659,-953){\makebox(0,0)[rb]{\smash{\SetFigFont{8}{9.6}{\familydefault}{\mddefault}{\updefault}0.6}}}
\put(659,-632){\makebox(0,0)[rb]{\smash{\SetFigFont{8}{9.6}{\familydefault}{\mddefault}{\updefault}0.7}}}
\put(659,-312){\makebox(0,0)[rb]{\smash{\SetFigFont{8}{9.6}{\familydefault}{\mddefault}{\updefault}0.8}}}
\put(659,  9){\makebox(0,0)[rb]{\smash{\SetFigFont{8}{9.6}{\familydefault}{\mddefault}{\updefault}0.9}}}
\put(659,329){\makebox(0,0)[rb]{\smash{\SetFigFont{8}{9.6}{\familydefault}{\mddefault}{\updefault}1}}}
\put(659,-2234){\makebox(0,0)[rb]{\smash{\SetFigFont{8}{9.6}{\familydefault}{\mddefault}{\updefault}0.2}}}
\put(733,-2999){\makebox(0,0)[b]{\smash{\SetFigFont{8}{9.6}{\familydefault}{\mddefault}{\updefault}0.6}}}
\put(659,-2875){\makebox(0,0)[rb]{\smash{\SetFigFont{8}{9.6}{\familydefault}{\mddefault}{\updefault}0}}}
\put(1498,-2999){\makebox(0,0)[b]{\smash{\SetFigFont{8}{9.6}{\familydefault}{\mddefault}{\updefault}0.8}}}
\put(2264,-2999){\makebox(0,0)[b]{\smash{\SetFigFont{8}{9.6}{\familydefault}{\mddefault}{\updefault}1}}}
\put(3029,-2999){\makebox(0,0)[b]{\smash{\SetFigFont{8}{9.6}{\familydefault}{\mddefault}{\updefault}1.2}}}
\put(3795,-2999){\makebox(0,0)[b]{\smash{\SetFigFont{8}{9.6}{\familydefault}{\mddefault}{\updefault}1.4}}}
\put(4560,-2999){\makebox(0,0)[b]{\smash{\SetFigFont{8}{9.6}{\familydefault}{\mddefault}{\updefault}1.6}}}
\put(5326,-2999){\makebox(0,0)[b]{\smash{\SetFigFont{8}{9.6}{\familydefault}{\mddefault}{\updefault}1.8}}}
\put(6091,-2999){\makebox(0,0)[b]{\smash{\SetFigFont{8}{9.6}{\familydefault}{\mddefault}{\updefault}2}}}
\put(659,-2555){\makebox(0,0)[rb]{\smash{\SetFigFont{8}{9.6}{\familydefault}{\mddefault}{\updefault}0.1}}}
\end{picture}
\caption{Plot of the average rod lengths in the isotropic (solid) and
nematic (dashed) shadow phases. The average rod length in the cloud
phases, being identical to that of the parent, is always one. Thus the
nematic phase always has a larger average length than its coexisting
isotropic phase. Fractionation disappears when the system becomes
monodisperse, $\sigma\to 0$, as expected.}
\label{fig:fractionation_unimodal}
\efig

\subsection{Bidisperse rod lengths}
\label{sec:bidisperse_res}

From Figures~\ref{fig:phase_diag_unimodal},
\ref{fig:vol_frac_unimodal} and~\ref{fig:fractionation_unimodal} we
saw that there is no I-N-N three-phase region and associated N-N
demixing within the present model, at least for a Schulz distribution
of lengths. On the other hand, previous work on bi- and tridisperse
systems~\cite{VroLek93,VroLek97} shows that such a three-phase
separation is possible within the full Onsager theory. The absence of
a three-phase region observed above could then be due to either the
difference in the length distributions investigated (continuous and
unimodal versus discrete with two or three different lengths), or to
our truncation of the Onsager theory that gave the $\Ptwo$ Onsager
model. To ascertain which of these applies, we now study the phase
diagram of a bidisperse system of rods within the $\Ptwo$ Onsager
model.
The normalized parent length distribution is then
\begin{equation}\label{eq:bidisperse_parent}
P^{(0)}(l)=\left[(1-x)\delta(l-l_1)+x\delta(l-l_2)\right]
\end{equation}
where $l_1<l_2$ and $x$ is the (number) fraction of longer rods. We
again fix the average length to unity, so that $l_1$ and $l_2$ can be
written in terms of their ratio $q=l_2/l_1$ as 
\begin{equation}
l_1=\frac{1}{1-x+xq}\label{eq:l1}, \qquad
l_2=ql_1\label{eq:l2}
\end{equation}
The polydispersity is given by
\begin{equation}\label{eq:bidisp_sigma}
\sigma^2 = \frac{x(1-x)(q-1)^2}{(1-x+xq)^2}
\end{equation}
and $\sigma$ becomes zero for $x=0$ and $x=1$, where we recover a
monodisperse system; its maximum for given $q$ is reached at
$x=1/(q+1)$.

\bfig
\begin{picture}(0,0)%
\epsfig{file=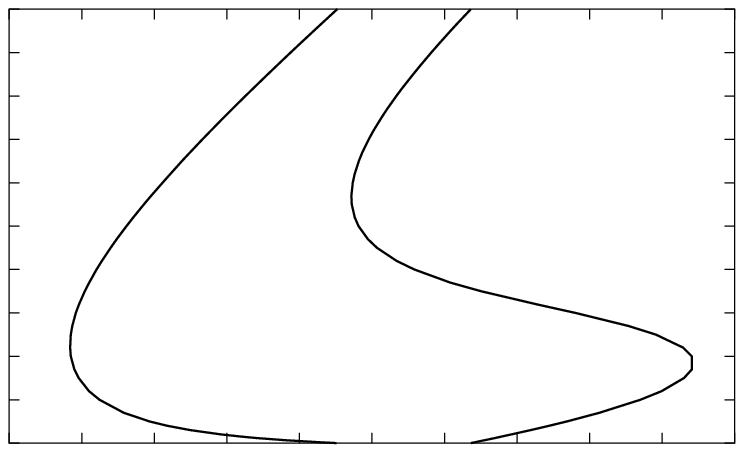}%
\end{picture}%
\setlength{\unitlength}{2565sp}%
\begingroup\makeatletter\ifx\SetFigFont\undefined%
\gdef\SetFigFont#1#2#3#4#5{%
  \reset@font\fontsize{#1}{#2pt}%
  \fontfamily{#3}\fontseries{#4}\fontshape{#5}%
  \selectfont}%
\fi\endgroup%
\begin{picture}(6145,3827)(76,-3361)
\put( 76,-1111){\makebox(0,0)[lb]{\smash{\SetFigFont{8}{9.6}{\familydefault}{\mddefault}{\updefault}$x$}}}
\put(3076,-3361){\makebox(0,0)[lb]{\smash{\SetFigFont{8}{9.6}{\familydefault}{\mddefault}{\updefault}$\rho_0$}}}
\put(4426,-1261){\makebox(0,0)[lb]{\smash{\SetFigFont{8}{9.6}{\familydefault}{\mddefault}{\updefault}$N$}}}
\put(1126,-811){\makebox(0,0)[lb]{\smash{\SetFigFont{8}{9.6}{\familydefault}{\mddefault}{\updefault}$I$}}}
\put(2176,-1936){\makebox(0,0)[lb]{\smash{\SetFigFont{8}{9.6}{\familydefault}{\mddefault}{\updefault}$I+N$}}}
\put(659,-953){\makebox(0,0)[rb]{\smash{\SetFigFont{8}{9.6}{\familydefault}{\mddefault}{\updefault}0.6}}}
\put(659,-632){\makebox(0,0)[rb]{\smash{\SetFigFont{8}{9.6}{\familydefault}{\mddefault}{\updefault}0.7}}}
\put(659,-312){\makebox(0,0)[rb]{\smash{\SetFigFont{8}{9.6}{\familydefault}{\mddefault}{\updefault}0.8}}}
\put(659,  9){\makebox(0,0)[rb]{\smash{\SetFigFont{8}{9.6}{\familydefault}{\mddefault}{\updefault}0.9}}}
\put(659,329){\makebox(0,0)[rb]{\smash{\SetFigFont{8}{9.6}{\familydefault}{\mddefault}{\updefault}1}}}
\put(733,-2999){\makebox(0,0)[b]{\smash{\SetFigFont{8}{9.6}{\familydefault}{\mddefault}{\updefault}2.6}}}
\put(1269,-2999){\makebox(0,0)[b]{\smash{\SetFigFont{8}{9.6}{\familydefault}{\mddefault}{\updefault}2.8}}}
\put(1805,-2999){\makebox(0,0)[b]{\smash{\SetFigFont{8}{9.6}{\familydefault}{\mddefault}{\updefault}3}}}
\put(2340,-2999){\makebox(0,0)[b]{\smash{\SetFigFont{8}{9.6}{\familydefault}{\mddefault}{\updefault}3.2}}}
\put(2876,-2999){\makebox(0,0)[b]{\smash{\SetFigFont{8}{9.6}{\familydefault}{\mddefault}{\updefault}3.4}}}
\put(659,-1273){\makebox(0,0)[rb]{\smash{\SetFigFont{8}{9.6}{\familydefault}{\mddefault}{\updefault}0.5}}}
\put(3412,-2999){\makebox(0,0)[b]{\smash{\SetFigFont{8}{9.6}{\familydefault}{\mddefault}{\updefault}3.6}}}
\put(659,-2875){\makebox(0,0)[rb]{\smash{\SetFigFont{8}{9.6}{\familydefault}{\mddefault}{\updefault}0}}}
\put(3948,-2999){\makebox(0,0)[b]{\smash{\SetFigFont{8}{9.6}{\familydefault}{\mddefault}{\updefault}3.8}}}
\put(4484,-2999){\makebox(0,0)[b]{\smash{\SetFigFont{8}{9.6}{\familydefault}{\mddefault}{\updefault}4}}}
\put(5019,-2999){\makebox(0,0)[b]{\smash{\SetFigFont{8}{9.6}{\familydefault}{\mddefault}{\updefault}4.2}}}
\put(5555,-2999){\makebox(0,0)[b]{\smash{\SetFigFont{8}{9.6}{\familydefault}{\mddefault}{\updefault}4.4}}}
\put(6091,-2999){\makebox(0,0)[b]{\smash{\SetFigFont{8}{9.6}{\familydefault}{\mddefault}{\updefault}4.6}}}
\put(659,-1593){\makebox(0,0)[rb]{\smash{\SetFigFont{8}{9.6}{\familydefault}{\mddefault}{\updefault}0.4}}}
\put(659,-1914){\makebox(0,0)[rb]{\smash{\SetFigFont{8}{9.6}{\familydefault}{\mddefault}{\updefault}0.3}}}
\put(659,-2234){\makebox(0,0)[rb]{\smash{\SetFigFont{8}{9.6}{\familydefault}{\mddefault}{\updefault}0.2}}}
\put(659,-2555){\makebox(0,0)[rb]{\smash{\SetFigFont{8}{9.6}{\familydefault}{\mddefault}{\updefault}0.1}}}
\end{picture}
\vspace*{0.2cm}
\caption{Phase diagram of a bidisperse system for rod length ratio
$q=2.5$. We only show the limits of the the I-N coexistence region,
which are given by the isotropic and nematic cloud curves, and omit
the shadow curves. All densities shown are therefore those of the
parent phase.}
\label{fig:bidisp_q_2_5}
\efig
In Fig.~\ref{fig:bidisp_q_2_5} we show the phase diagram~\cite{bidisperse_note}
for rod length ratio $q=2.5$. 
The density of the parent at the transition from
single-phase to two-phase regions is plotted against the number
fraction of longer rods on the vertical axis; effectively, we are only
showing the cloud curves and omitting the shadow curves.  This ``phase
boundaries only'' representation will be useful below when three-phase
equilibria appear (which are represented on the cloud and shadow
curves only in terms of the exceptional points where direct
transitions from one to three phases occur). Comparing with the Schulz
distribution case in Fig.~\ref{fig:phase_diag_unimodal}, where the
solid lines correspond to those of Fig.~\ref{fig:bidisp_q_2_5}, we see
that again the coexistence region broadens as we move from either of
the monodisperse limits $x=0$ and $x=1$ towards larger polydispersity.

For comparison with previous results on bidisperse systems obtained
from the full Onsager theory~\cite{BuiLek93,VroLek93}, it will also be
useful to show our phase diagrams in the representation employed in
Refs.~\onlinecite{BuiLek93,VroLek93}, which uses the variables
$\tilde\phi_1=(L_1/D)\phi_1$ and $\tilde\phi_2=(L_1/D)\phi_2$ instead
of our $x$ and $\rho_0$. Here $\phi_1$ and $\phi_2$ are the volume
fractions of short and long rods, and $L_1=L_0 l_1$ is the
unnormalized length of the short rods. With $N_1$ and $N_2$ the number
of short and long rods, we have
\begin{equation}
\tilde\phi_i=
\frac{L_1}{D}\phi_i = \frac{L_1}{D}\frac{\pi}{4}D^2 L_i \frac{N_i}{V}
= \frac{N_i}{N} l_1 l_i \left(\frac{\pi}{4} DL_0^2 \frac{N}{V}\right)
\end{equation}
The term in brackets is just our dimensionless rod number density
$\rho_0$, so that
\begin{eqnarray}
\tilde\phi_1&=&(1-x)l_1^2\rho_0\label{eq:2phi1}\\
\tilde\phi_2&=&xl_1 l_2\rho_0\label{eq:2phi2}\
\end{eqnarray}
This gives us the relation between the two representations of the
phase diagram. As an aside, we note that
our convention for choosing the normalized rod lengths $l_1=L_1/L_0$
and $l_2=L_2/L_0$ differs from that of Ref.~\onlinecite{VroLek93}; we vary $l_1$ and
$l_2$ according to Eq.~\eqref{eq:l1} to maintain a constant average length
in the parent, while in Ref.~\onlinecite{VroLek93} constant values of $l_1$ and
$l_2$ are used. However, these assignments can alternatively be
thought of as arising from a different choice of the reference length
$L_0$ for identical values of $L_1$ and $L_2$. Since $L_0$ does not
enter the definitions~\eqq{eq:2phi1}{eq:2phi2}, the values of the
$\tilde\phi_1$ and $\tilde\phi_2$ remain
unaffected. We show the $\tilde\phi_1$, $\tilde\phi_2$-representation
for the case $q=2.5$ in 
Fig.~\ref{fig:bidisp_vroege_q_2_5}.
\bfig
\begin{picture}(0,0)%
\epsfig{file=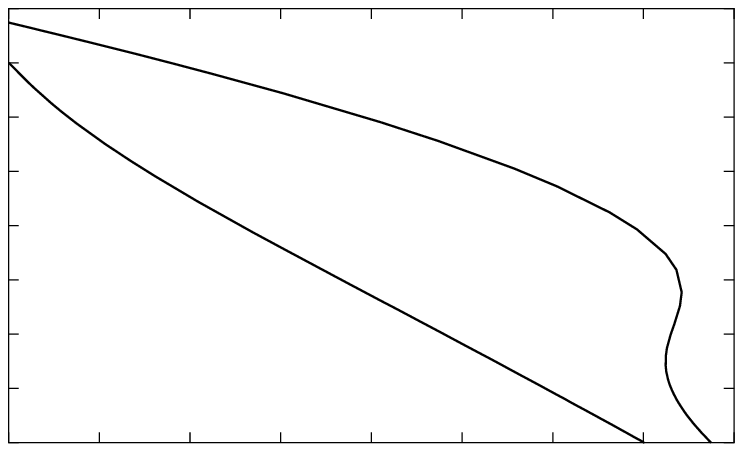}%
\end{picture}%
\setlength{\unitlength}{2565sp}%
\begingroup\makeatletter\ifx\SetFigFont\undefined%
\gdef\SetFigFont#1#2#3#4#5{%
  \reset@font\fontsize{#1}{#2pt}%
  \fontfamily{#3}\fontseries{#4}\fontshape{#5}%
  \selectfont}%
\fi\endgroup%
\begin{picture}(6295,3806)(-74,-3340)
\put(3076,-1111){\makebox(0,0)[lb]{\smash{\SetFigFont{8}{9.6}{\familydefault}{\mddefault}{\updefault}$I+N$}}}
\put(5101,-511){\makebox(0,0)[lb]{\smash{\SetFigFont{8}{9.6}{\familydefault}{\mddefault}{\updefault}$N$}}}
\put(2026,-1936){\makebox(0,0)[lb]{\smash{\SetFigFont{8}{9.6}{\familydefault}{\mddefault}{\updefault}$I$}}}
\put(-74,-1036){\makebox(0,0)[lb]{\smash{\SetFigFont{8}{9.6}{\familydefault}{\mddefault}{\updefault}$\tilde\phi_1$}}}
\put(3151,-3286){\makebox(0,0)[lb]{\smash{\SetFigFont{8}{9.6}{\familydefault}{\mddefault}{\updefault}$\tilde\phi_2$}}}
\put(659,-472){\makebox(0,0)[rb]{\smash{\SetFigFont{8}{9.6}{\familydefault}{\mddefault}{\updefault}3}}}
\put(659,-71){\makebox(0,0)[rb]{\smash{\SetFigFont{8}{9.6}{\familydefault}{\mddefault}{\updefault}3.5}}}
\put(659,329){\makebox(0,0)[rb]{\smash{\SetFigFont{8}{9.6}{\familydefault}{\mddefault}{\updefault}4}}}
\put(733,-2999){\makebox(0,0)[b]{\smash{\SetFigFont{8}{9.6}{\familydefault}{\mddefault}{\updefault}0}}}
\put(1403,-2999){\makebox(0,0)[b]{\smash{\SetFigFont{8}{9.6}{\familydefault}{\mddefault}{\updefault}0.2}}}
\put(2073,-2999){\makebox(0,0)[b]{\smash{\SetFigFont{8}{9.6}{\familydefault}{\mddefault}{\updefault}0.4}}}
\put(2742,-2999){\makebox(0,0)[b]{\smash{\SetFigFont{8}{9.6}{\familydefault}{\mddefault}{\updefault}0.6}}}
\put(3412,-2999){\makebox(0,0)[b]{\smash{\SetFigFont{8}{9.6}{\familydefault}{\mddefault}{\updefault}0.8}}}
\put(659,-872){\makebox(0,0)[rb]{\smash{\SetFigFont{8}{9.6}{\familydefault}{\mddefault}{\updefault}2.5}}}
\put(4082,-2999){\makebox(0,0)[b]{\smash{\SetFigFont{8}{9.6}{\familydefault}{\mddefault}{\updefault}1}}}
\put(659,-2875){\makebox(0,0)[rb]{\smash{\SetFigFont{8}{9.6}{\familydefault}{\mddefault}{\updefault}0}}}
\put(4752,-2999){\makebox(0,0)[b]{\smash{\SetFigFont{8}{9.6}{\familydefault}{\mddefault}{\updefault}1.2}}}
\put(5421,-2999){\makebox(0,0)[b]{\smash{\SetFigFont{8}{9.6}{\familydefault}{\mddefault}{\updefault}1.4}}}
\put(6091,-2999){\makebox(0,0)[b]{\smash{\SetFigFont{8}{9.6}{\familydefault}{\mddefault}{\updefault}1.6}}}
\put(659,-1273){\makebox(0,0)[rb]{\smash{\SetFigFont{8}{9.6}{\familydefault}{\mddefault}{\updefault}2}}}
\put(659,-1673){\makebox(0,0)[rb]{\smash{\SetFigFont{8}{9.6}{\familydefault}{\mddefault}{\updefault}1.5}}}
\put(659,-2074){\makebox(0,0)[rb]{\smash{\SetFigFont{8}{9.6}{\familydefault}{\mddefault}{\updefault}1}}}
\put(659,-2474){\makebox(0,0)[rb]{\smash{\SetFigFont{8}{9.6}{\familydefault}{\mddefault}{\updefault}0.5}}}
\end{picture}
\caption{Phase diagram of a bidisperse system with $q=2.5$ in terms of
the rescaled volume fractions $\tilde\phi_1$ and $\tilde\phi_2$ of
long and short rods, respectively. This representation is useful to
compare with previous results for bidisperse systems from the full
Onsager theory; it also shows directly at any point of the phase
diagram the composition of the parent phase.}
\label{fig:bidisp_vroege_q_2_5}
\efig

By including Legendre polynomials up to ${\mathcal P}_{14}$, and
determining the orientational distribution functions $P_l(\theta)$
numerically, Lekkerkerker {\em et al.}~\cite{LekCouVanDeb84} obtained
good approximations to the phase diagrams for the full Onsager theory
of bidisperse rods. They found that at $q=5$ there is a re-entrant
nematic phase, while at $q=2$ there is no such re-entrance. Buining
and Lekkerkerker~\cite{BuiLek93} also found no re-entrance at $q=2.5$;
the latter 
finding agrees with our above results. Using a Gaussian
variational approximation for the $P_l(\theta)$, Odijk and
Lekkerkerker~\cite{OdiLek85} showed that re-entrance is
generically to be expected for large $q$. Vroege and
Lekkerkerker~\cite{VroLek93} later found, using the same approach,
that in addition to the re-entrance one obtains three-phase I-N-N and
two-phase N-N equilibria
for $q$ above $\approx 3.17$. Their approximation predicted that the
boundaries of the N-N coexistence region should be independent of the rod
number density $\rho_0$, corresponding to horizontal lines in a
($\rho_0$,$x$) representation of the phase diagram or to lines through
the origin in a ($\tilde\phi_1$,$\tilde\phi_2$) plot.
Van Roij and Mulder~\cite{VanMul96} later showed that
these statements are exactly true (only) in the limit of large
$\rho_0$; this means in particular that the N-N coexistence region is
not closed off by an N-N critical point. Birshtein et
al.~\cite{BirKolPry88}, using a trial function of the 
type used by Onsager~\cite{Onsager49}, had earlier come to the
opposite conclusion. However, they emphasized (as did Abe and
Flory~\cite{AbeFlo78}) the difficulties of solving the phase 
equilibrium conditions at large density, which reduced the reliability
of their results.

To compare the above findings from treatments of the full Onsager
theory with the $\Ptwo$ Onsager model, we now investigate how the
phase diagram changes with increasing $q$.
\bfig %\input{bidisp_q_3_5_both}
\begin{picture}(0,0)%
\epsfig{file=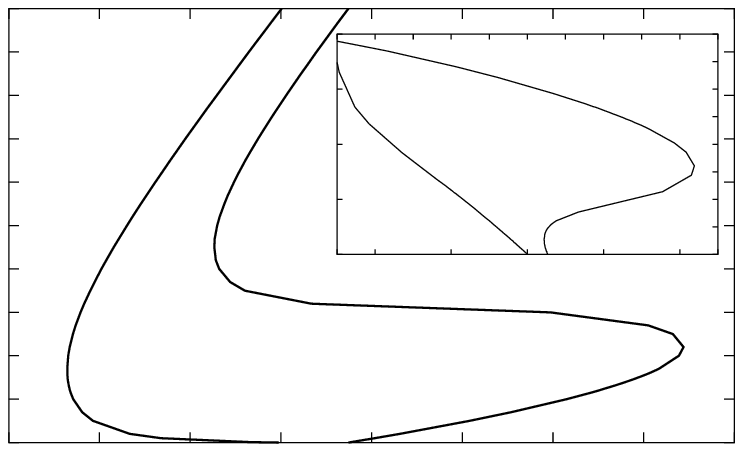}%
\end{picture}%
\setlength{\unitlength}{2565sp}%
\begingroup\makeatletter\ifx\SetFigFont\undefined%
\gdef\SetFigFont#1#2#3#4#5{%
  \reset@font\fontsize{#1}{#2pt}%
  \fontfamily{#3}\fontseries{#4}\fontshape{#5}%
  \selectfont}%
\fi\endgroup%
\begin{picture}(5990,3752)(151,-3286)
\put(5293,-179){\makebox(0,0)[lb]{\smash{\SetFigFont{8}{9.6}{\familydefault}{\mddefault}{\updefault}$N$}}}
\put(4308,-713){\makebox(0,0)[lb]{\smash{\SetFigFont{8}{9.6}{\familydefault}{\mddefault}{\updefault}$I+N$}}}
\put(3481,-1207){\makebox(0,0)[lb]{\smash{\SetFigFont{8}{9.6}{\familydefault}{\mddefault}{\updefault}$I$}}}
\put(5251,-1786){\makebox(0,0)[lb]{\smash{\SetFigFont{8}{9.6}{\familydefault}{\mddefault}{\updefault}$\tilde\phi_2$}}}
\put(3076,-3286){\makebox(0,0)[lb]{\smash{\SetFigFont{8}{9.6}{\familydefault}{\mddefault}{\updefault}$\rho_0$}}}
\put(2626,-811){\makebox(0,0)[lb]{\smash{\SetFigFont{8}{9.6}{\familydefault}{\mddefault}{\updefault}$\tilde\phi_1$}}}
\put(1201,-1186){\makebox(0,0)[lb]{\smash{\SetFigFont{8}{9.6}{\familydefault}{\mddefault}{\updefault}$I$}}}
\put(5326,-2611){\makebox(0,0)[lb]{\smash{\SetFigFont{8}{9.6}{\familydefault}{\mddefault}{\updefault}$N$}}}
\put(151,-811){\makebox(0,0)[lb]{\smash{\SetFigFont{8}{9.6}{\familydefault}{\mddefault}{\updefault}$x$}}}
\put(2326,-2311){\makebox(0,0)[lb]{\smash{\SetFigFont{8}{9.6}{\familydefault}{\mddefault}{\updefault}$I+N$}}}
\put(733,-2999){\makebox(0,0)[b]{\smash{\SetFigFont{8}{9.6}{\familydefault}{\mddefault}{\updefault}2}}}
\put(1403,-2999){\makebox(0,0)[b]{\smash{\SetFigFont{8}{9.6}{\familydefault}{\mddefault}{\updefault}2.5}}}
\put(2073,-2999){\makebox(0,0)[b]{\smash{\SetFigFont{8}{9.6}{\familydefault}{\mddefault}{\updefault}3}}}
\put(2742,-2999){\makebox(0,0)[b]{\smash{\SetFigFont{8}{9.6}{\familydefault}{\mddefault}{\updefault}3.5}}}
\put(3412,-2999){\makebox(0,0)[b]{\smash{\SetFigFont{8}{9.6}{\familydefault}{\mddefault}{\updefault}4}}}
\put(4082,-2999){\makebox(0,0)[b]{\smash{\SetFigFont{8}{9.6}{\familydefault}{\mddefault}{\updefault}4.5}}}
\put(4752,-2999){\makebox(0,0)[b]{\smash{\SetFigFont{8}{9.6}{\familydefault}{\mddefault}{\updefault}5}}}
\put(5421,-2999){\makebox(0,0)[b]{\smash{\SetFigFont{8}{9.6}{\familydefault}{\mddefault}{\updefault}5.5}}}
\put(6091,-2999){\makebox(0,0)[b]{\smash{\SetFigFont{8}{9.6}{\familydefault}{\mddefault}{\updefault}6}}}
\put(659,  9){\makebox(0,0)[rb]{\smash{\SetFigFont{8}{9.6}{\familydefault}{\mddefault}{\updefault}0.9}}}
\put(659,-312){\makebox(0,0)[rb]{\smash{\SetFigFont{8}{9.6}{\familydefault}{\mddefault}{\updefault}0.8}}}
\put(659,-632){\makebox(0,0)[rb]{\smash{\SetFigFont{8}{9.6}{\familydefault}{\mddefault}{\updefault}0.7}}}
\put(659,-953){\makebox(0,0)[rb]{\smash{\SetFigFont{8}{9.6}{\familydefault}{\mddefault}{\updefault}0.6}}}
\put(659,-1273){\makebox(0,0)[rb]{\smash{\SetFigFont{8}{9.6}{\familydefault}{\mddefault}{\updefault}0.5}}}
\put(659,329){\makebox(0,0)[rb]{\smash{\SetFigFont{8}{9.6}{\familydefault}{\mddefault}{\updefault}1}}}
\put(659,-1593){\makebox(0,0)[rb]{\smash{\SetFigFont{8}{9.6}{\familydefault}{\mddefault}{\updefault}0.4}}}
\put(659,-2875){\makebox(0,0)[rb]{\smash{\SetFigFont{8}{9.6}{\familydefault}{\mddefault}{\updefault}0}}}
\put(3117,-1455){\makebox(0,0)[rb]{\smash{\SetFigFont{8}{9.6}{\familydefault}{\mddefault}{\updefault}0}}}
\put(3117,-1049){\makebox(0,0)[rb]{\smash{\SetFigFont{8}{9.6}{\familydefault}{\mddefault}{\updefault}1}}}
\put(3117,-642){\makebox(0,0)[rb]{\smash{\SetFigFont{8}{9.6}{\familydefault}{\mddefault}{\updefault}2}}}
\put(3117,-236){\makebox(0,0)[rb]{\smash{\SetFigFont{8}{9.6}{\familydefault}{\mddefault}{\updefault}3}}}
\put(3117,170){\makebox(0,0)[rb]{\smash{\SetFigFont{8}{9.6}{\familydefault}{\mddefault}{\updefault}4}}}
\put(659,-1914){\makebox(0,0)[rb]{\smash{\SetFigFont{8}{9.6}{\familydefault}{\mddefault}{\updefault}0.3}}}
\put(659,-2234){\makebox(0,0)[rb]{\smash{\SetFigFont{8}{9.6}{\familydefault}{\mddefault}{\updefault}0.2}}}
\put(659,-2555){\makebox(0,0)[rb]{\smash{\SetFigFont{8}{9.6}{\familydefault}{\mddefault}{\updefault}0.1}}}
\put(3438,-1586){\makebox(0,0)[b]{\smash{\SetFigFont{8}{9.6}{\familydefault}{\mddefault}{\updefault}0.2}}}
\put(4000,-1586){\makebox(0,0)[b]{\smash{\SetFigFont{8}{9.6}{\familydefault}{\mddefault}{\updefault}0.6}}}
\put(4563,-1586){\makebox(0,0)[b]{\smash{\SetFigFont{8}{9.6}{\familydefault}{\mddefault}{\updefault}1}}}
\put(5126,-1586){\makebox(0,0)[b]{\smash{\SetFigFont{8}{9.6}{\familydefault}{\mddefault}{\updefault}1.4}}}
\put(5689,-1586){\makebox(0,0)[b]{\smash{\SetFigFont{8}{9.6}{\familydefault}{\mddefault}{\updefault}1.8}}}
\end{picture}
\vspace*{0.1cm}
\caption{Phase diagram of a bidisperse system with $q=3.5$,
represented in terms of the parent density $\rho_0$ and the long
rod number fraction $x$ (main plot) and in terms of the rescaled volume
fractions $\tilde\phi_1$ and $\tilde\phi_2$ (inset).  The large
``bump'' of the nematic cloud point curve (towards large densities in
the main figure) narrowly avoids a re-entrance of the nematic phase.}
\label{fig:bidisp_q_3_5_both} 
\efig
Figs.~\ref{fig:bidisp_q_3_5_both}, \ref{fig:bidisp_q_7_both},
\ref{fig:bidisperse_phase_diag} and \ref{fig:bidisperse_vroegelike}
show the phase diagrams for rod length ratios $q=3.5$, $q=7$ and
$q=12$ in both representations ($\rho_0,x$ and
$\tilde\phi_1,\tilde\phi_2$). At $q=3.5$ a re-entrance of the nematic
phase has almost appeared; at $q=7$ this is fully developed. At
$q=12$ (Fig.~\ref{fig:bidisperse_phase_diag}), finally, we have in
addition a three-phase I-N-N region 
bordered by a region (N-N) of coexistence of two nematics~\cite{phase_diag}. The N-N
region is closed off by an N-N critical point, where the moments
$\rho_1$ and $\rho_2$ in the two nematics become identical.
These features
persist at higher $q$; for $q=30$, for example, the phase diagram (not
shown) has the same structure as Fig.~\ref{fig:bidisperse_phase_diag}
but with both the re-entrance and the three-phase region being located
at smaller $x$ and extending to higher densities. For even larger $q$
the numerical evaluation of the phase diagrams becomes more
troublesome because the interesting features of the phase diagram
shift towards smaller and smaller $x$ where even with the moment
method the phase equilibrium conditions cannot be reliably
solved numerically. However, a simple approximate treatment (see
App.~\ref{sec:SNO_approx}) can be used to make progress and
understand how the phase diagram scales for large $q$.
\bfig
\begin{picture}(0,0)%
\epsfig{file=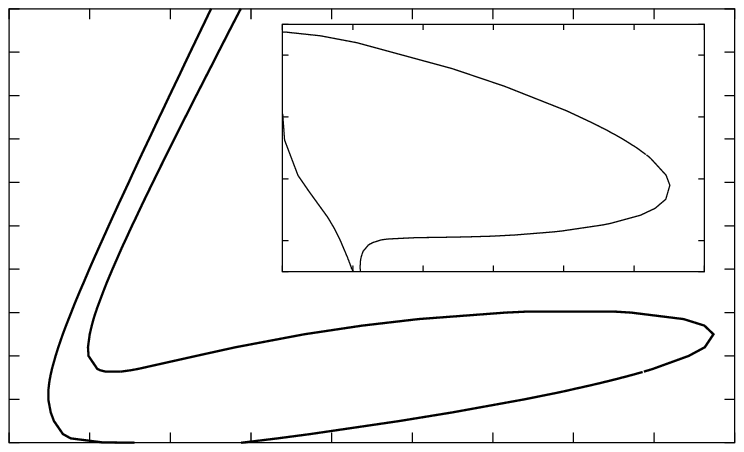}%
\end{picture}%
\setlength{\unitlength}{2565sp}%
\begingroup\makeatletter\ifx\SetFigFont\undefined%
\gdef\SetFigFont#1#2#3#4#5{%
  \reset@font\fontsize{#1}{#2pt}%
  \fontfamily{#3}\fontseries{#4}\fontshape{#5}%
  \selectfont}%
\fi\endgroup%
\begin{picture}(5890,3752)(526,-3511)
\put(2401,-811){\makebox(0,0)[lb]{\smash{\SetFigFont{8}{9.6}{\familydefault}{\mddefault}{\updefault}$\tilde\phi_1$}}}
\put(5701,-2011){\makebox(0,0)[lb]{\smash{\SetFigFont{8}{9.6}{\familydefault}{\mddefault}{\updefault}$\tilde\phi_2$}}}
\put(5603,-633){\makebox(0,0)[lb]{\smash{\SetFigFont{8}{9.6}{\familydefault}{\mddefault}{\updefault}$N$}}}
\put(3816,-1060){\makebox(0,0)[lb]{\smash{\SetFigFont{8}{9.6}{\familydefault}{\mddefault}{\updefault}$I+N$}}}
\put(3226,-3511){\makebox(0,0)[lb]{\smash{\SetFigFont{8}{9.6}{\familydefault}{\mddefault}{\updefault}$\rho_0$}}}
\put(3118,-1487){\makebox(0,0)[lb]{\smash{\SetFigFont{8}{9.6}{\familydefault}{\mddefault}{\updefault}$I$}}}
\put(1201,-1711){\makebox(0,0)[lb]{\smash{\SetFigFont{8}{9.6}{\familydefault}{\mddefault}{\updefault}$I$}}}
\put(5401,-2911){\makebox(0,0)[lb]{\smash{\SetFigFont{8}{9.6}{\familydefault}{\mddefault}{\updefault}$N$}}}
\put(526,-1036){\makebox(0,0)[lb]{\smash{\SetFigFont{8}{9.6}{\familydefault}{\mddefault}{\updefault}$x$}}}
\put(2101,-2761){\makebox(0,0)[lb]{\smash{\SetFigFont{8}{9.6}{\familydefault}{\mddefault}{\updefault}$I+N$}}}
\put(958,-3224){\makebox(0,0)[b]{\smash{\SetFigFont{8}{9.6}{\familydefault}{\mddefault}{\updefault}1}}}
\put(1553,-3224){\makebox(0,0)[b]{\smash{\SetFigFont{8}{9.6}{\familydefault}{\mddefault}{\updefault}2}}}
\put(2149,-3224){\makebox(0,0)[b]{\smash{\SetFigFont{8}{9.6}{\familydefault}{\mddefault}{\updefault}3}}}
\put(3339,-3224){\makebox(0,0)[b]{\smash{\SetFigFont{8}{9.6}{\familydefault}{\mddefault}{\updefault}5}}}
\put(3935,-3224){\makebox(0,0)[b]{\smash{\SetFigFont{8}{9.6}{\familydefault}{\mddefault}{\updefault}6}}}
\put(4530,-3224){\makebox(0,0)[b]{\smash{\SetFigFont{8}{9.6}{\familydefault}{\mddefault}{\updefault}7}}}
\put(5125,-3224){\makebox(0,0)[b]{\smash{\SetFigFont{8}{9.6}{\familydefault}{\mddefault}{\updefault}8}}}
\put(5721,-3224){\makebox(0,0)[b]{\smash{\SetFigFont{8}{9.6}{\familydefault}{\mddefault}{\updefault}9}}}
\put(6316,-3224){\makebox(0,0)[b]{\smash{\SetFigFont{8}{9.6}{\familydefault}{\mddefault}{\updefault}10}}}
\put(884,-216){\makebox(0,0)[rb]{\smash{\SetFigFont{8}{9.6}{\familydefault}{\mddefault}{\updefault}0.9}}}
\put(884,-537){\makebox(0,0)[rb]{\smash{\SetFigFont{8}{9.6}{\familydefault}{\mddefault}{\updefault}0.8}}}
\put(884,-857){\makebox(0,0)[rb]{\smash{\SetFigFont{8}{9.6}{\familydefault}{\mddefault}{\updefault}0.7}}}
\put(884,-1178){\makebox(0,0)[rb]{\smash{\SetFigFont{8}{9.6}{\familydefault}{\mddefault}{\updefault}0.6}}}
\put(2934,-1580){\makebox(0,0)[rb]{\smash{\SetFigFont{8}{9.6}{\familydefault}{\mddefault}{\updefault}0.5}}}
\put(2934,-1124){\makebox(0,0)[rb]{\smash{\SetFigFont{8}{9.6}{\familydefault}{\mddefault}{\updefault}1.5}}}
\put(884,104){\makebox(0,0)[rb]{\smash{\SetFigFont{8}{9.6}{\familydefault}{\mddefault}{\updefault}1}}}
\put(2934,-667){\makebox(0,0)[rb]{\smash{\SetFigFont{8}{9.6}{\familydefault}{\mddefault}{\updefault}2.5}}}
\put(884,-3100){\makebox(0,0)[rb]{\smash{\SetFigFont{8}{9.6}{\familydefault}{\mddefault}{\updefault}0}}}
\put(5573,-1901){\makebox(0,0)[b]{\smash{\SetFigFont{8}{9.6}{\familydefault}{\mddefault}{\updefault}2.5}}}
\put(884,-1498){\makebox(0,0)[rb]{\smash{\SetFigFont{8}{9.6}{\familydefault}{\mddefault}{\updefault}0.5}}}
\put(884,-1818){\makebox(0,0)[rb]{\smash{\SetFigFont{8}{9.6}{\familydefault}{\mddefault}{\updefault}0.4}}}
\put(884,-2139){\makebox(0,0)[rb]{\smash{\SetFigFont{8}{9.6}{\familydefault}{\mddefault}{\updefault}0.3}}}
\put(884,-2459){\makebox(0,0)[rb]{\smash{\SetFigFont{8}{9.6}{\familydefault}{\mddefault}{\updefault}0.2}}}
\put(884,-2780){\makebox(0,0)[rb]{\smash{\SetFigFont{8}{9.6}{\familydefault}{\mddefault}{\updefault}0.1}}}
\put(2744,-3224){\makebox(0,0)[b]{\smash{\SetFigFont{8}{9.6}{\familydefault}{\mddefault}{\updefault}4}}}
\put(2934,-211){\makebox(0,0)[rb]{\smash{\SetFigFont{8}{9.6}{\familydefault}{\mddefault}{\updefault}3.5}}}
\put(4015,-1901){\makebox(0,0)[b]{\smash{\SetFigFont{8}{9.6}{\familydefault}{\mddefault}{\updefault}1}}}
\put(3526,-1901){\makebox(0,0)[b]{\smash{\SetFigFont{8}{9.6}{\familydefault}{\mddefault}{\updefault}0.5}}}
\put(3001,-1901){\makebox(0,0)[rb]{\smash{\SetFigFont{8}{9.6}{\familydefault}{\mddefault}{\updefault}0}}}
\put(4534,-1901){\makebox(0,0)[b]{\smash{\SetFigFont{8}{9.6}{\familydefault}{\mddefault}{\updefault}1.5}}}
\put(5053,-1901){\makebox(0,0)[b]{\smash{\SetFigFont{8}{9.6}{\familydefault}{\mddefault}{\updefault}2}}}
\put(6092,-1901){\makebox(0,0)[b]{\smash{\SetFigFont{8}{9.6}{\familydefault}{\mddefault}{\updefault}3}}}
\end{picture}
\vspace*{0.1cm}
\caption{Phase diagram of a bidisperse system with $q=7$ in both
the ($\rho_0,x$) and the ($\tilde\phi_2,\tilde\phi_1$)
representations. The re-entrance of the nematic phase is now fully
developed, but there is not yet an N-N or three-phase (I-N-N) region.
}
\label{fig:bidisp_q_7_both}
\efig
\bfig
\begin{picture}(0,0)%
\epsfig{file=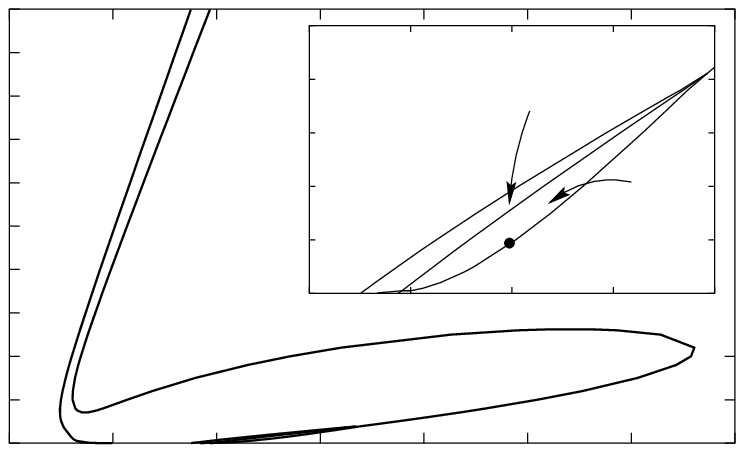}%
\end{picture}%
\setlength{\unitlength}{2565sp}%
\begingroup\makeatletter\ifx\SetFigFont\undefined%
\gdef\SetFigFont#1#2#3#4#5{%
  \reset@font\fontsize{#1}{#2pt}%
  \fontfamily{#3}\fontseries{#4}\fontshape{#5}%
  \selectfont}%
\fi\endgroup%
\begin{picture}(5965,3827)(226,-3361)
\put(2776,-3361){\makebox(0,0)[lb]{\smash{\SetFigFont{8}{9.6}{\familydefault}{\mddefault}{\updefault}$\rho_0$}}}
\put(5176,-1336){\makebox(0,0)[lb]{\smash{\SetFigFont{8}{9.6}{\familydefault}{\mddefault}{\updefault}$N$}}}
\put(5101,-1036){\makebox(0,0)[lb]{\smash{\SetFigFont{8}{9.6}{\familydefault}{\mddefault}{\updefault}$N_1+N_2$}}}
\put(4276,-286){\makebox(0,0)[lb]{\smash{\SetFigFont{8}{9.6}{\familydefault}{\mddefault}{\updefault}$I+N_1+N_2$}}}
\put(2176,-2611){\makebox(0,0)[lb]{\smash{\SetFigFont{8}{9.6}{\familydefault}{\mddefault}{\updefault}$I+N$}}}
\put(4276,-1561){\makebox(0,0)[lb]{\smash{\SetFigFont{8}{9.6}{\familydefault}{\mddefault}{\updefault}critical point}}}
\put(3376,-811){\makebox(0,0)[lb]{\smash{\SetFigFont{8}{9.6}{\familydefault}{\mddefault}{\updefault}$I+N$}}}
\put(226,-511){\makebox(0,0)[lb]{\smash{\SetFigFont{8}{9.6}{\familydefault}{\mddefault}{\updefault}$x$}}}
\put(4576,-2761){\makebox(0,0)[lb]{\smash{\SetFigFont{8}{9.6}{\familydefault}{\mddefault}{\updefault}$N$}}}
\put(901,-1861){\makebox(0,0)[lb]{\smash{\SetFigFont{8}{9.6}{\familydefault}{\mddefault}{\updefault}$I$}}}
\put(733,-2999){\makebox(0,0)[b]{\smash{\SetFigFont{8}{9.6}{\familydefault}{\mddefault}{\updefault}0}}}
\put(1498,-2999){\makebox(0,0)[b]{\smash{\SetFigFont{8}{9.6}{\familydefault}{\mddefault}{\updefault}2}}}
\put(2264,-2999){\makebox(0,0)[b]{\smash{\SetFigFont{8}{9.6}{\familydefault}{\mddefault}{\updefault}4}}}
\put(3029,-2999){\makebox(0,0)[b]{\smash{\SetFigFont{8}{9.6}{\familydefault}{\mddefault}{\updefault}6}}}
\put(3795,-2999){\makebox(0,0)[b]{\smash{\SetFigFont{8}{9.6}{\familydefault}{\mddefault}{\updefault}8}}}
\put(4560,-2999){\makebox(0,0)[b]{\smash{\SetFigFont{8}{9.6}{\familydefault}{\mddefault}{\updefault}10}}}
\put(5326,-2999){\makebox(0,0)[b]{\smash{\SetFigFont{8}{9.6}{\familydefault}{\mddefault}{\updefault}12}}}
\put(6091,-2999){\makebox(0,0)[b]{\smash{\SetFigFont{8}{9.6}{\familydefault}{\mddefault}{\updefault}14}}}
\put(659,  9){\makebox(0,0)[rb]{\smash{\SetFigFont{8}{9.6}{\familydefault}{\mddefault}{\updefault}0.9}}}
\put(659,-312){\makebox(0,0)[rb]{\smash{\SetFigFont{8}{9.6}{\familydefault}{\mddefault}{\updefault}0.8}}}
\put(659,-632){\makebox(0,0)[rb]{\smash{\SetFigFont{8}{9.6}{\familydefault}{\mddefault}{\updefault}0.7}}}
\put(659,-953){\makebox(0,0)[rb]{\smash{\SetFigFont{8}{9.6}{\familydefault}{\mddefault}{\updefault}0.6}}}
\put(2906,-1746){\makebox(0,0)[rb]{\smash{\SetFigFont{8}{9.6}{\familydefault}{\mddefault}{\updefault}0}}}
\put(2906,-1351){\makebox(0,0)[rb]{\smash{\SetFigFont{8}{9.6}{\familydefault}{\mddefault}{\updefault}0.01}}}
\put(659,329){\makebox(0,0)[rb]{\smash{\SetFigFont{8}{9.6}{\familydefault}{\mddefault}{\updefault}1}}}
\put(2906,-956){\makebox(0,0)[rb]{\smash{\SetFigFont{8}{9.6}{\familydefault}{\mddefault}{\updefault}0.02}}}
\put(659,-2875){\makebox(0,0)[rb]{\smash{\SetFigFont{8}{9.6}{\familydefault}{\mddefault}{\updefault}0}}}
\put(2906,-560){\makebox(0,0)[rb]{\smash{\SetFigFont{8}{9.6}{\familydefault}{\mddefault}{\updefault}0.03}}}
\put(2906,-165){\makebox(0,0)[rb]{\smash{\SetFigFont{8}{9.6}{\familydefault}{\mddefault}{\updefault}0.04}}}
\put(2906,230){\makebox(0,0)[rb]{\smash{\SetFigFont{8}{9.6}{\familydefault}{\mddefault}{\updefault}0.05}}}
\put(3697,-1851){\makebox(0,0)[b]{\smash{\SetFigFont{8}{9.6}{\familydefault}{\mddefault}{\updefault}4}}}
\put(5193,-1822){\makebox(0,0)[b]{\smash{\SetFigFont{8}{9.6}{\familydefault}{\mddefault}{\updefault}6}}}
\put(659,-1273){\makebox(0,0)[rb]{\smash{\SetFigFont{8}{9.6}{\familydefault}{\mddefault}{\updefault}0.5}}}
\put(659,-1593){\makebox(0,0)[rb]{\smash{\SetFigFont{8}{9.6}{\familydefault}{\mddefault}{\updefault}0.4}}}
\put(659,-1914){\makebox(0,0)[rb]{\smash{\SetFigFont{8}{9.6}{\familydefault}{\mddefault}{\updefault}0.3}}}
\put(659,-2234){\makebox(0,0)[rb]{\smash{\SetFigFont{8}{9.6}{\familydefault}{\mddefault}{\updefault}0.2}}}
\put(2948,-1851){\makebox(0,0)[b]{\smash{\SetFigFont{8}{9.6}{\familydefault}{\mddefault}{\updefault}3}}}
\put(4444,-1851){\makebox(0,0)[b]{\smash{\SetFigFont{8}{9.6}{\familydefault}{\mddefault}{\updefault}5}}}
\put(5941,-1851){\makebox(0,0)[b]{\smash{\SetFigFont{8}{9.6}{\familydefault}{\mddefault}{\updefault}7}}}
\put(659,-2555){\makebox(0,0)[rb]{\smash{\SetFigFont{8}{9.6}{\familydefault}{\mddefault}{\updefault}0.1}}}
\end{picture}
\vspace*{0,1cm}
\caption{Phase diagram of a bidisperse system with $q=12$, in terms of
the parent density $\rho_0$  and the long rod number fraction $x$.
As shown in detail in the inset, a three-phase I-N-N region appears,
bordered by a region of N-N coexistence; at larger values of $x$, the
nematic phase is re-entrant.}
\label{fig:bidisperse_phase_diag}
\efig 

In summary of our results for the bidisperse case, an increase in $q$
produces similar qualitative changes in the structure of the phase
diagram as for the full Onsager theory. The quantitative details of
course differ, with the onset of both the nematic re-entrance and the
three-phase region shifted to higher $q$. The only qualitative
difference caused by the fact that we are using the truncated $\Ptwo$
Onsager model is that, in contrast to the full Onsager
theory~\cite{VanMul96}, the N-N region
does not extend to large densities but is instead closed by a critical
point, whose properties we discuss further in
App.~\ref{sec:SNO_approx}. 
\bfig
\begin{picture}(0,0)%
\epsfig{file=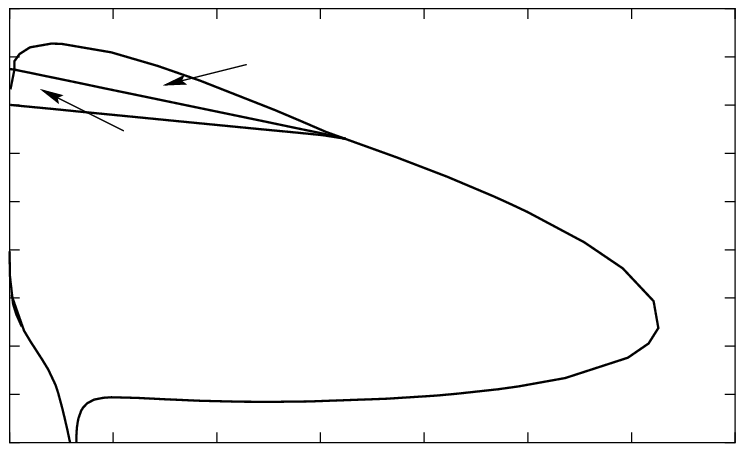}%
\end{picture}%
\setlength{\unitlength}{2565sp}%
\begingroup\makeatletter\ifx\SetFigFont\undefined%
\gdef\SetFigFont#1#2#3#4#5{%
  \reset@font\fontsize{#1}{#2pt}%
  \fontfamily{#3}\fontseries{#4}\fontshape{#5}%
  \selectfont}%
\fi\endgroup%
\begin{picture}(6220,3806)(301,-3190)
\put(2476,239){\makebox(0,0)[lb]{\smash{\SetFigFont{8}{9.6}{\familydefault}{\mddefault}{\updefault}$N_1+N_2$}}}
\put(1726,-586){\makebox(0,0)[lb]{\smash{\SetFigFont{8}{9.6}{\familydefault}{\mddefault}{\updefault}$I+N_1+N_2$}}}
\put(1051,-2461){\makebox(0,0)[lb]{\smash{\SetFigFont{8}{9.6}{\familydefault}{\mddefault}{\updefault}$I$}}}
\put(5401,-361){\makebox(0,0)[lb]{\smash{\SetFigFont{8}{9.6}{\familydefault}{\mddefault}{\updefault}$N$}}}
\put(3376,-3136){\makebox(0,0)[lb]{\smash{\SetFigFont{8}{9.6}{\familydefault}{\mddefault}{\updefault}$\tilde\phi_2$}}}
\put(301,-736){\makebox(0,0)[lb]{\smash{\SetFigFont{8}{9.6}{\familydefault}{\mddefault}{\updefault}$\tilde\phi_1$}}}
\put(2401,-1336){\makebox(0,0)[lb]{\smash{\SetFigFont{8}{9.6}{\familydefault}{\mddefault}{\updefault}$I+N$}}}
\put(959,123){\makebox(0,0)[rb]{\smash{\SetFigFont{8}{9.6}{\familydefault}{\mddefault}{\updefault}4}}}
\put(959,479){\makebox(0,0)[rb]{\smash{\SetFigFont{8}{9.6}{\familydefault}{\mddefault}{\updefault}4.5}}}
\put(1033,-2849){\makebox(0,0)[b]{\smash{\SetFigFont{8}{9.6}{\familydefault}{\mddefault}{\updefault}0}}}
\put(1798,-2849){\makebox(0,0)[b]{\smash{\SetFigFont{8}{9.6}{\familydefault}{\mddefault}{\updefault}0.5}}}
\put(2564,-2849){\makebox(0,0)[b]{\smash{\SetFigFont{8}{9.6}{\familydefault}{\mddefault}{\updefault}1}}}
\put(3329,-2849){\makebox(0,0)[b]{\smash{\SetFigFont{8}{9.6}{\familydefault}{\mddefault}{\updefault}1.5}}}
\put(4095,-2849){\makebox(0,0)[b]{\smash{\SetFigFont{8}{9.6}{\familydefault}{\mddefault}{\updefault}2}}}
\put(4860,-2849){\makebox(0,0)[b]{\smash{\SetFigFont{8}{9.6}{\familydefault}{\mddefault}{\updefault}2.5}}}
\put(959,-233){\makebox(0,0)[rb]{\smash{\SetFigFont{8}{9.6}{\familydefault}{\mddefault}{\updefault}3.5}}}
\put(5626,-2849){\makebox(0,0)[b]{\smash{\SetFigFont{8}{9.6}{\familydefault}{\mddefault}{\updefault}3}}}
\put(959,-2725){\makebox(0,0)[rb]{\smash{\SetFigFont{8}{9.6}{\familydefault}{\mddefault}{\updefault}0}}}
\put(6391,-2849){\makebox(0,0)[b]{\smash{\SetFigFont{8}{9.6}{\familydefault}{\mddefault}{\updefault}3.5}}}
\put(959,-589){\makebox(0,0)[rb]{\smash{\SetFigFont{8}{9.6}{\familydefault}{\mddefault}{\updefault}3}}}
\put(959,-945){\makebox(0,0)[rb]{\smash{\SetFigFont{8}{9.6}{\familydefault}{\mddefault}{\updefault}2.5}}}
\put(959,-1301){\makebox(0,0)[rb]{\smash{\SetFigFont{8}{9.6}{\familydefault}{\mddefault}{\updefault}2}}}
\put(959,-1657){\makebox(0,0)[rb]{\smash{\SetFigFont{8}{9.6}{\familydefault}{\mddefault}{\updefault}1.5}}}
\put(959,-2013){\makebox(0,0)[rb]{\smash{\SetFigFont{8}{9.6}{\familydefault}{\mddefault}{\updefault}1}}}
\put(959,-2369){\makebox(0,0)[rb]{\smash{\SetFigFont{8}{9.6}{\familydefault}{\mddefault}{\updefault}0.5}}}
\end{picture}
\caption{Phase diagram of a bidisperse system with $q=12$ in terms of
$\tilde\phi_1$ and $\tilde\phi_2$, the scaled volume fractions of
shorter and longer rods at $q=12$.}
\label{fig:bidisperse_vroegelike}
\efig  
This distinction between the $\Ptwo$ Onsager model and the full
Onsager theory can be made plausible by considering the effect of our
truncation on the excluded volume term, i.e., the excess free energy,
in the limit of large
densities $\rho_0$ where nematic phases will be strongly ordered. In
the full Onsager theory, it can be shown~\cite{VanMul96,VroLek93} that
the excess free energy density is simply $\fexc=2\rho_0$ for large
$\rho_0$ so that the excluded volume per particle is
$\fexc/\rho_0=2$. This simple scaling with $\rho_0$ arises because the
typical rod angles $\theta$ are $\sim 1/\rho_0$ for large $\rho_0$;
the angular part $K(\theta,\theta')$ of the excluded volume
interaction, as given by Eq.~\eqref{eq:kernel}, scales {\em linearly}
with the typical values of $\theta$ and $\theta'$ and is thus also
proportional to $1/\rho_0$. The high density phase behaviour thus
arises solely from a competition between the two components of the
ideal part of the free energy~\eqref{eq:free_en}, the entropy of
mixing and the orientational entropy; N-N demixing occurs when the
resulting gain in orientational entropy dominates the loss of entropy
of mixing. In our $\Ptwo$ truncation, on the other hand, $K$ is
approximated as $K(\theta,\theta') \sim c_1 - c_2
\Ptwo(\cos\theta)\Ptwo(\cos\theta^\prime)$ which becomes $\sim a + b[\theta^2
+ (\theta')^2]$ for small angles, with constants $a$ and $b$. For
large $\rho_0$, when the typical values of $\theta$ and $\theta'$
again become small, this approaches a constant rather than decrease to
zero as in the full Onsager theory, giving a quadratic density scaling
of the excess free energy. The same
conclusion can be reached from the moment description: for $\theta\to
0$ we have $\rho_2=\rho_1$ and the excess free energy density becomes
$\frac{1}{2}(c_1-c_2)\rho_1^2$. The quadratic scaling of the excess
free energy with density, as opposed to the linear scaling in the full
Onsager theory, will act as a driving force against phase separation.
This is in line with our findings above regarding the absence of N-N
demixing at high density. One can in fact show formally (see
App.~\ref{sec:SNO_approx}) that N-N coexistence is not possible in the
$\Ptwo$ Onsager model in the limit of large $\rho_0$.

The above argument can in fact be extended to all possible truncations
of the full Onsager theory, since the inclusion of any finite number
of higher order Legendre polynomials will still give a behaviour
$K(\theta,\theta') \sim a + b[\theta^2 + (\theta')^2]$ of the
orientational part of the excluded volume for small angles; the
correct linear scaling with $\theta$ and $\theta'$ of the full theory
is obtained only when all terms in the infinite series are
retained. Equivalently, in the moment description one would have a
finite number of moments analogous to $\rho_2$, but these would all
saturate to values proportional to $\rho_1$ at high density,
giving again the quadratic scaling $\fexc\sim\rho_1^2$.
We conclude, therefore, that there cannot be any N-N region at
sufficiently high density for any truncation of the full Onsager
theory; this is consistent with van Roij and Mulder's
explanation~\cite{VanMul96} for Birshtein's
result of an N-N critical point in his approximate treatment of
bidisperse Onsager theory. Of course, for any finite density we must
eventually recover the correct phase behaviour as more and more higher
order Legendre polynomials are included to give an increasingly good
representation of $K(\theta,\theta')$ for nonzero angles. One would
expect, therefore, that successively higher order truncations of
Onsager theory would lead to N-N coexistence regions extending to
higher and higher densities, and eventually diverging as the order of
truncation is taken to infinity.

Before passing to
a different length distribution, let us note that in the volume
fraction representation (e.g.\ Fig.~\ref{fig:bidisperse_vroegelike}) of the
phase diagrams of our bidisperse system all tielines connecting
coexisting phases are necessarily straight. This implies, in
particular, that the 
boundaries of the three phase region, which are formed by two-phase
tielines, are straight. The same argument no longer applies in the
scenario that we study next: when more than two different rod lengths
are present, the two-dimensional phase diagrams are cuts through a
higher-dimensional phase diagrams, and generic tielines no longer lie
within this cut plane. 

\subsection{Bimodal length distribution}
\label{sec:bimodal_res}

Our results for the bidisperse case suggest that the $\Ptwo$ Onsager
model correctly reproduces most of the qualitative features of the
phase diagram of the full Onsager theory, except for N-N coexistence
at high densities. This suggests similar qualitative agreement also in
the case of unimodal length distributions. We therefore conclude that
the differences between unimodal and bidisperse distributions that we
found, in particular the absence of re-entrant features and N-N and
I-N-N phase equilibria in the unimodal case, are physical and not due
to our truncation of the model. To understand in more detail how these
features arise in dependence on the shape of the length distribution,
we now turn to a family of length distributions that lets us
interpolate between the two limits of bidisperse and unimodal
distributions. The normalized parent length distributions we consider
are mixtures of two Schulz distributions, of the form
\begin{equation}\label{eq:bimodal_parent}
P^{(0)}(l)=(1-x)l_1^{-1}S\left({l}/{l_1}\right)+x
l_2^{-1}S\left({l}/{l_2}\right)
\end{equation}
Here $S(l)$ denotes the Schulz distribution~\eqref{eq:schulz} and
$l_1$ and $l_2$ are again given by Eq.~\eqref{eq:l1} to constrain the
average rod length to unity. Let us call $\ssch$ the normalized width
(standard deviation) of the Schulz distribution, to distinguish it
from the normalized overall standard deviation $\sigma$ of the parent
distribution; the distribution~\eqref{eq:bimodal_parent} is then
characterized by the three parameters $q$, $x$ and $\ssch$. For
$\ssch\rightarrow 0$, we recover the bidisperse case; on the other
hand, as $\ssch$ increases for given $x$ and $q$, the distribution
eventually becomes unimodal as shown in
Fig.~\ref{fig:bimodal_parent_both}. For given $q$ and $x$, the value of
$\ssch$ at which this happens is determined by the condition that
there must be a value of $l$ for which $P^{(0)}(l)$ has vanishing
first and second derivative; the resulting set of equations can easily
be solved numerically for $\ssch$.
\bfig
\begin{picture}(0,0)%
\epsfig{file=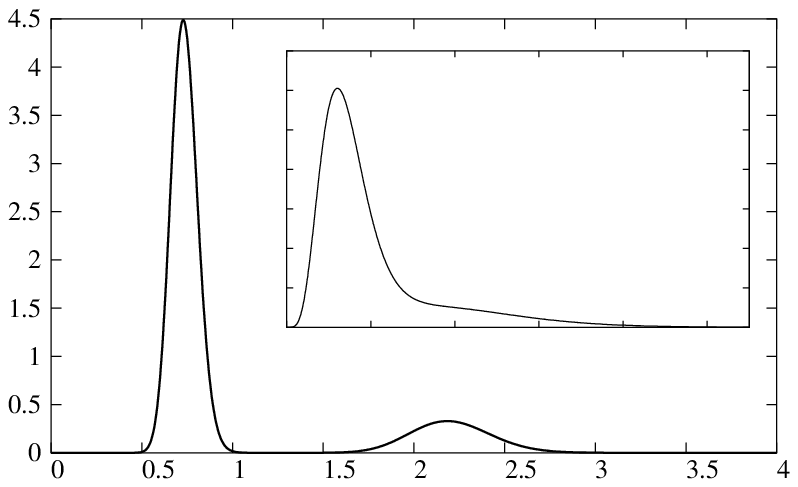}%
\end{picture}%
\setlength{\unitlength}{2565sp}%
\begingroup\makeatletter\ifx\SetFigFont\undefined%
\gdef\SetFigFont#1#2#3#4#5{%
  \reset@font\fontsize{#1}{#2pt}%
  \fontfamily{#3}\fontseries{#4}\fontshape{#5}%
  \selectfont}%
\fi\endgroup%
\begin{picture}(6037,3762)(76,-3211)
\put(2926,-3211){\makebox(0,0)[lb]{\smash{\SetFigFont{8}{9.6}{\familydefault}{\mddefault}{\updefault}$l$}}}
\put( 76,-811){\makebox(0,0)[lb]{\smash{\SetFigFont{8}{9.6}{\familydefault}{\mddefault}{\updefault}$P^{(0)}$}}}
\put(4014,-2113){\makebox(0,0)[lb]{\smash{\SetFigFont{8}{9.6}{\familydefault}{\mddefault}{\updefault}$l$}}}
\put(2425,-976){\makebox(0,0)[rb]{\smash{\SetFigFont{8}{9.6}{\familydefault}{\mddefault}{\updefault}0.6}}}
\put(2425,-684){\makebox(0,0)[rb]{\smash{\SetFigFont{8}{9.6}{\familydefault}{\mddefault}{\updefault}0.8}}}
\put(2425,-392){\makebox(0,0)[rb]{\smash{\SetFigFont{8}{9.6}{\familydefault}{\mddefault}{\updefault}1}}}
\put(2425,-101){\makebox(0,0)[rb]{\smash{\SetFigFont{8}{9.6}{\familydefault}{\mddefault}{\updefault}1.2}}}
\put(2425,191){\makebox(0,0)[rb]{\smash{\SetFigFont{8}{9.6}{\familydefault}{\mddefault}{\updefault}1.4}}}
\put(2425,-1559){\makebox(0,0)[rb]{\smash{\SetFigFont{8}{9.6}{\familydefault}{\mddefault}{\updefault}0.2}}}
\put(2425,-1268){\makebox(0,0)[rb]{\smash{\SetFigFont{8}{9.6}{\familydefault}{\mddefault}{\updefault}0.4}}}
\put(2425,-1851){\makebox(0,0)[rb]{\smash{\SetFigFont{8}{9.6}{\familydefault}{\mddefault}{\updefault}0}}}
\put(5577,-1971){\makebox(0,0)[b]{\smash{\SetFigFont{8}{9.6}{\familydefault}{\mddefault}{\updefault}5}}}
\put(2472,-1971){\makebox(0,0)[b]{\smash{\SetFigFont{8}{9.6}{\familydefault}{\mddefault}{\updefault}0}}}
\put(3093,-1971){\makebox(0,0)[b]{\smash{\SetFigFont{8}{9.6}{\familydefault}{\mddefault}{\updefault}1}}}
\put(3714,-1971){\makebox(0,0)[b]{\smash{\SetFigFont{8}{9.6}{\familydefault}{\mddefault}{\updefault}2}}}
\put(4335,-1971){\makebox(0,0)[b]{\smash{\SetFigFont{8}{9.6}{\familydefault}{\mddefault}{\updefault}3}}}
\put(4956,-1971){\makebox(0,0)[b]{\smash{\SetFigFont{8}{9.6}{\familydefault}{\mddefault}{\updefault}4}}}
\end{picture}
\caption{Examples of our bimodal parent length distribution. In the limit
where the width of each of the two peaks tends to zero, the bidisperse
case is recovered. As the width is increased, on the other hand, the
distribution eventually becomes unimodal; in the inset, we show the
case where the minimum and the second maximum have just merged into a
turning point.}
\label{fig:bimodal_parent_both}
\efig

Our aim is now to understand how the occurrence of the more exotic
features of the phase diagram, such as an I-N-N three-phase region or
a re-entrant nematic phase, depends on the rod length distribution;
one might, for example, suspect a dependence on bimodality (the
presence of two separate peaks in the distribution) or on the overall
polydispersity.

We will proceed in two separate ways. First, we fix a value of $q$
(=12) for which in the bidisperse case ($\ssch=0$) we have a
three-phase region and re-entrance in the nematic phase. We then
increase $\ssch$ and monitor the corresponding variation of the
($\rho_0,x$) phase diagram. When $\ssch$ becomes sufficiently large,
we have a unimodal length distribution and thus expect to see
behaviour more akin to that for the Schulz length distribution. For
fixed $q$, the three-phase region should thus disappear when $\ssch$
reaches a sufficiently large value. The process will essentially be
the opposite of what we have done so far: in the previous sections we
began from a unimodal phase diagram and moved to the bidisperse case
with larger and larger rod length ratio $q$. Now we start from the
largest value of $q$ that we used for a bidisperse distribution, and
widen the two peaks of the distribution until we return to a unimodal
(though not a simple Schulz) distribution.

The second procedure will be to fix values of $x$ and $q$ for which we
have coexistence of three phases in the bidisperse limit $\ssch\to 0$,
and to start increasing the polydispersity. This lets us check
whether the existence of a three-phase region is linked directly to the
bimodality of the parent, by comparing the value of $\ssch$ for which
the region of three-phase equilibrium disappears to that where the
parent length distribution changes from bimodal to unimodal. 

Adopting the first procedure, we have found the phase diagrams for a
small normalized width $\ssch=0.1$ of the two peaks in the bimodal
length distribution. For $q=3.5$ the result is a phase diagram that is
practically identical to Fig.~\ref{fig:bidisp_q_3_5_both}; the system
is still effectively bidisperse. More interestingly, the same
conclusion applies when we introduce a small peak width for $q=12$:
with $\ssch=0.1$, we again get phase diagrams that are essentially
indistinguishale from the bidisperse case shown in
Figs.~\ref{fig:bidisperse_phase_diag}
and~\ref{fig:bidisperse_vroegelike}. In particular, neither the
three-phase region nor the re-entrant nematic phase are significantly
affected by such a small degree of polydispersity of the two peaks.
Larger values of $\ssch$, on the other hand, do have a significant
effect. If the width of the peaks is increased to $\ssch=0.6$, for
example (Fig.~\ref{fig:phased_q_12_sigma_0.6}), the phase diagram
changes noticeably. In particular, the three-phase region has now
disappeared, or at least shrunk to values below $x=10^{-3}$ where our
numerical calculations become unreliable, and the nematic re-entrance
has become much less pronounced. This is in line with the intuition
that for increasing $\ssch$ behaviour similar to that for the unimodal
Schulz distribution should be recovered.
\bfig
\begin{picture}(0,0)%
\epsfig{file=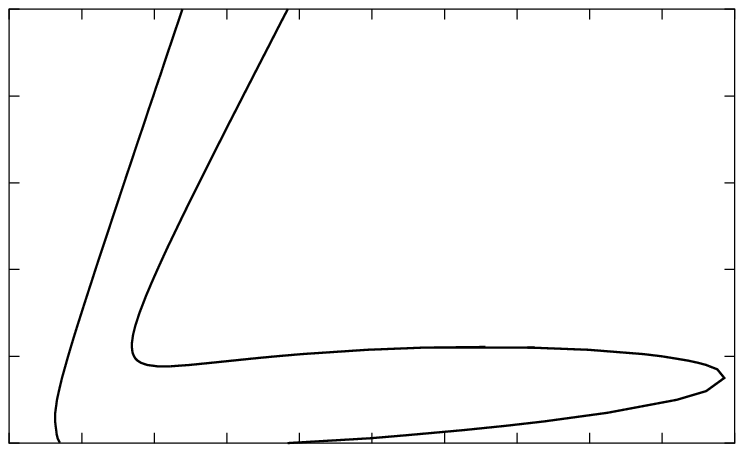}%
\end{picture}%
\setlength{\unitlength}{2565sp}%
\begingroup\makeatletter\ifx\SetFigFont\undefined%
\gdef\SetFigFont#1#2#3#4#5{%
  \reset@font\fontsize{#1}{#2pt}%
  \fontfamily{#3}\fontseries{#4}\fontshape{#5}%
  \selectfont}%
\fi\endgroup%
\begin{picture}(6115,3827)(76,-3361)
\put(2701,-1561){\makebox(0,0)[lb]{\smash{\SetFigFont{8}{9.6}{\familydefault}{\mddefault}{\updefault}$N$}}}
\put(3001,-3361){\makebox(0,0)[lb]{\smash{\SetFigFont{8}{9.6}{\familydefault}{\mddefault}{\updefault}$\rho_0$}}}
\put(2476,-2611){\makebox(0,0)[lb]{\smash{\SetFigFont{8}{9.6}{\familydefault}{\mddefault}{\updefault}$I+N$}}}
\put( 76,-1336){\makebox(0,0)[lb]{\smash{\SetFigFont{8}{9.6}{\familydefault}{\mddefault}{\updefault}$x$}}}
\put(1051,-1111){\makebox(0,0)[lb]{\smash{\SetFigFont{8}{9.6}{\familydefault}{\mddefault}{\updefault}$I$}}}
\put(733,-2999){\makebox(0,0)[b]{\smash{\SetFigFont{8}{9.6}{\familydefault}{\mddefault}{\updefault}0}}}
\put(1269,-2999){\makebox(0,0)[b]{\smash{\SetFigFont{8}{9.6}{\familydefault}{\mddefault}{\updefault}1}}}
\put(1805,-2999){\makebox(0,0)[b]{\smash{\SetFigFont{8}{9.6}{\familydefault}{\mddefault}{\updefault}2}}}
\put(2340,-2999){\makebox(0,0)[b]{\smash{\SetFigFont{8}{9.6}{\familydefault}{\mddefault}{\updefault}3}}}
\put(2876,-2999){\makebox(0,0)[b]{\smash{\SetFigFont{8}{9.6}{\familydefault}{\mddefault}{\updefault}4}}}
\put(3412,-2999){\makebox(0,0)[b]{\smash{\SetFigFont{8}{9.6}{\familydefault}{\mddefault}{\updefault}5}}}
\put(3948,-2999){\makebox(0,0)[b]{\smash{\SetFigFont{8}{9.6}{\familydefault}{\mddefault}{\updefault}6}}}
\put(659,329){\makebox(0,0)[rb]{\smash{\SetFigFont{8}{9.6}{\familydefault}{\mddefault}{\updefault}1}}}
\put(4484,-2999){\makebox(0,0)[b]{\smash{\SetFigFont{8}{9.6}{\familydefault}{\mddefault}{\updefault}7}}}
\put(659,-2875){\makebox(0,0)[rb]{\smash{\SetFigFont{8}{9.6}{\familydefault}{\mddefault}{\updefault}0}}}
\put(5019,-2999){\makebox(0,0)[b]{\smash{\SetFigFont{8}{9.6}{\familydefault}{\mddefault}{\updefault}8}}}
\put(5555,-2999){\makebox(0,0)[b]{\smash{\SetFigFont{8}{9.6}{\familydefault}{\mddefault}{\updefault}9}}}
\put(6091,-2999){\makebox(0,0)[b]{\smash{\SetFigFont{8}{9.6}{\familydefault}{\mddefault}{\updefault}10}}}
\put(659,-312){\makebox(0,0)[rb]{\smash{\SetFigFont{8}{9.6}{\familydefault}{\mddefault}{\updefault}0.8}}}
\put(659,-953){\makebox(0,0)[rb]{\smash{\SetFigFont{8}{9.6}{\familydefault}{\mddefault}{\updefault}0.6}}}
\put(659,-1593){\makebox(0,0)[rb]{\smash{\SetFigFont{8}{9.6}{\familydefault}{\mddefault}{\updefault}0.4}}}
\put(659,-2234){\makebox(0,0)[rb]{\smash{\SetFigFont{8}{9.6}{\familydefault}{\mddefault}{\updefault}0.2}}}
\end{picture}
\caption{Phase diagram for a bimodal length distribution with rod
length ratio $q=12$ and normalized width of the two peaks of
$\ssch=0.6$. Compared to the bidisperse case with the same $q$
(Fig.~\ref{fig:bidisperse_phase_diag}), the three-phase (I-N-N) region
as well as the associated N-N region have disappeared. The re-entrance
of the nematic phase is still present, but is now visibly less
pronounced than in the bidisperse case or for small $\ssch$, e.g.\
$\ssch=0.1$. Towards $x=0$ and $x=1$ the coexistence region is now also rather
broader, since for the limiting values of $x$ we now recover a system
with a Schulz distribution of rod lengths, rather than a monodisperse
system.}
\label{fig:phased_q_12_sigma_0.6}
\efig
The results so far confirm that, in order for a three-phase region to
exist in the phase diagram, the two peaks of the parent length
distribution must, in some sense, be sufficiently well
``separated''. This notion is supported by an analysis of the large
$q$ limit in App.~\ref{sec:SNO_approx}. There we show that, for large
$q$ and values of $x$ of order $1/q$, there is an N-N region and a
corresponding three-phase region for all $\ssch$, up to the maximum
value $\ssch$ that we consider; correspondingly, one easily sees that
in this limit the length distribution always has two separate maxima.

We now proceed to a closer investigation of the link between the
presence of three-phase equilibria and bimodality of the length
distribution, using our second approach explained above:
we fix $q$ and $x$ so that the three-phase coexistence occurs
in the bidisperse limit $\ssch=0$, and then find the value of
$\ssch$ at which the three-phase region disappears, comparing with the
value where the parent changes from bimodal to unimodal.

Figs.~\ref{fig:bimodal_sigma_0_015} and~\ref{fig:bimodal_sigma_0_03}
show for $q=12$ and two different values of $x$ that the three-phase
region indeed disappears on increasing $\ssch$ above a certain value.
This threshold value of $\ssch$ is smaller for larger $x$, showing
that in the ($\rho_0$,$x$) phase diagram the top corner of the
three-phase region shrinks towards smaller values of $x$ as $\ssch$ is
increased. The $x$-value marking the bottom end of the three-phase
region will also depend on $\ssch$. One would suppose that it
increases with $\ssch$ until the $x$-values for the top and bottom
eventually coincide and the three-phase region disappears. However,
since in the phase diagrams we considered the lower end of the
three-phase region 
is located at extremely small $x$-values we were not able to confirm this
explicitly. 
\bfig
\begin{picture}(0,0)%
\epsfig{file=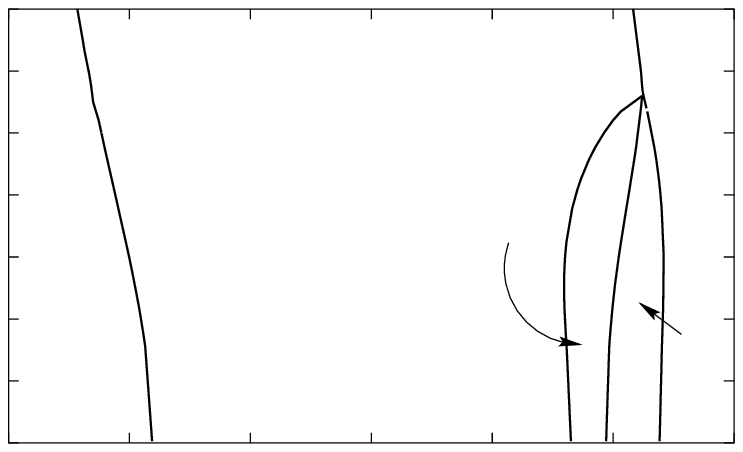}%
\end{picture}%
\setlength{\unitlength}{2565sp}%
\begingroup\makeatletter\ifx\SetFigFont\undefined%
\gdef\SetFigFont#1#2#3#4#5{%
  \reset@font\fontsize{#1}{#2pt}%
  \fontfamily{#3}\fontseries{#4}\fontshape{#5}%
  \selectfont}%
\fi\endgroup%
\begin{picture}(6215,3677)(-74,-3211)
\put(5251,-2236){\makebox(0,0)[lb]{\smash{\SetFigFont{8}{9.6}{\familydefault}{\mddefault}{\updefault}$N_1+N_2$}}}
\put(3976,-1186){\makebox(0,0)[lb]{\smash{\SetFigFont{8}{9.6}{\familydefault}{\mddefault}{\updefault}$I+N_1+N_2$}}}
\put(2926,-3211){\makebox(0,0)[lb]{\smash{\SetFigFont{8}{9.6}{\familydefault}{\mddefault}{\updefault}$\rho_0$ }}}
\put(2476,-1186){\makebox(0,0)[lb]{\smash{\SetFigFont{8}{9.6}{\familydefault}{\mddefault}{\updefault}$I+N$}}}
\put(-74,-811){\makebox(0,0)[lb]{\smash{\SetFigFont{8}{9.6}{\familydefault}{\mddefault}{\updefault}$\sigma_S$}}}
\put(1051,-1561){\makebox(0,0)[lb]{\smash{\SetFigFont{8}{9.6}{\familydefault}{\mddefault}{\updefault}$I$}}}
\put(5626,-661){\makebox(0,0)[lb]{\smash{\SetFigFont{8}{9.6}{\familydefault}{\mddefault}{\updefault}$N$}}}
\put(733,-2999){\makebox(0,0)[b]{\smash{\SetFigFont{8}{9.6}{\familydefault}{\mddefault}{\updefault}0}}}
\put(1626,-2999){\makebox(0,0)[b]{\smash{\SetFigFont{8}{9.6}{\familydefault}{\mddefault}{\updefault}1}}}
\put(2519,-2999){\makebox(0,0)[b]{\smash{\SetFigFont{8}{9.6}{\familydefault}{\mddefault}{\updefault}2}}}
\put(3412,-2999){\makebox(0,0)[b]{\smash{\SetFigFont{8}{9.6}{\familydefault}{\mddefault}{\updefault}3}}}
\put(4305,-2999){\makebox(0,0)[b]{\smash{\SetFigFont{8}{9.6}{\familydefault}{\mddefault}{\updefault}4}}}
\put(5198,-2999){\makebox(0,0)[b]{\smash{\SetFigFont{8}{9.6}{\familydefault}{\mddefault}{\updefault}5}}}
\put(659,329){\makebox(0,0)[rb]{\smash{\SetFigFont{8}{9.6}{\familydefault}{\mddefault}{\updefault}0.7}}}
\put(6091,-2999){\makebox(0,0)[b]{\smash{\SetFigFont{8}{9.6}{\familydefault}{\mddefault}{\updefault}6}}}
\put(659,-2875){\makebox(0,0)[rb]{\smash{\SetFigFont{8}{9.6}{\familydefault}{\mddefault}{\updefault}0}}}
\put(659,-129){\makebox(0,0)[rb]{\smash{\SetFigFont{8}{9.6}{\familydefault}{\mddefault}{\updefault}0.6}}}
\put(659,-586){\makebox(0,0)[rb]{\smash{\SetFigFont{8}{9.6}{\familydefault}{\mddefault}{\updefault}0.5}}}
\put(659,-1044){\makebox(0,0)[rb]{\smash{\SetFigFont{8}{9.6}{\familydefault}{\mddefault}{\updefault}0.4}}}
\put(659,-1502){\makebox(0,0)[rb]{\smash{\SetFigFont{8}{9.6}{\familydefault}{\mddefault}{\updefault}0.3}}}
\put(659,-1960){\makebox(0,0)[rb]{\smash{\SetFigFont{8}{9.6}{\familydefault}{\mddefault}{\updefault}0.2}}}
\put(659,-2417){\makebox(0,0)[rb]{\smash{\SetFigFont{8}{9.6}{\familydefault}{\mddefault}{\updefault}0.1}}}
\end{picture}
\caption{Phase behaviour of a bimodal system with $q=12$, $x=0.015$.
Shown is the density $\rho_0$ of the parent at which phase
transitions occur, against (on the vertical axis) the width
$\ssch$ of the two peaks of the bimodal parent.}
\label{fig:bimodal_sigma_0_015}
\efig
\bfig
\begin{picture}(0,0)%
\epsfig{file=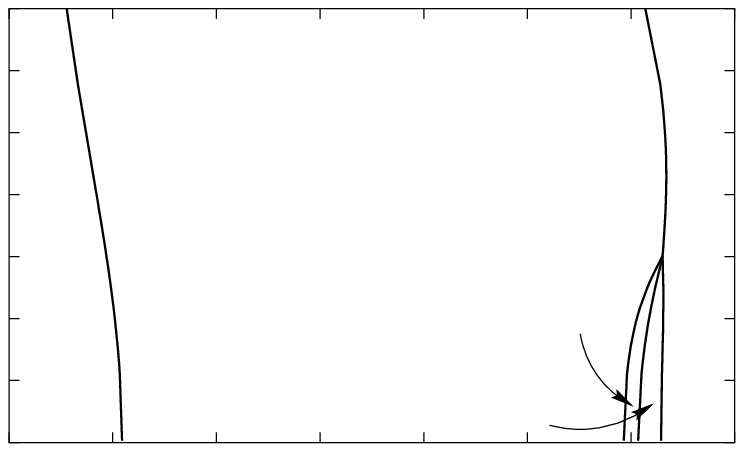}%
\end{picture}%
\setlength{\unitlength}{2565sp}%
\begingroup\makeatletter\ifx\SetFigFont\undefined%
\gdef\SetFigFont#1#2#3#4#5{%
  \reset@font\fontsize{#1}{#2pt}%
  \fontfamily{#3}\fontseries{#4}\fontshape{#5}%
  \selectfont}%
\fi\endgroup%
\begin{picture}(6140,3752)(1,-3286)
\put(1051,-1711){\makebox(0,0)[lb]{\smash{\SetFigFont{8}{9.6}{\familydefault}{\mddefault}{\updefault}$I$}}}
\put(4126,-2536){\makebox(0,0)[lb]{\smash{\SetFigFont{8}{9.6}{\familydefault}{\mddefault}{\updefault}$N_1+N_2$}}}
\put(4126,-1936){\makebox(0,0)[lb]{\smash{\SetFigFont{8}{9.6}{\familydefault}{\mddefault}{\updefault}$I+N_1+N_2$}}}
\put(2701,-1786){\makebox(0,0)[lb]{\smash{\SetFigFont{8}{9.6}{\familydefault}{\mddefault}{\updefault}$I+N$}}}
\put(2926,-3286){\makebox(0,0)[lb]{\smash{\SetFigFont{8}{9.6}{\familydefault}{\mddefault}{\updefault}$\rho_0$}}}
\put(5776,-1561){\makebox(0,0)[lb]{\smash{\SetFigFont{8}{9.6}{\familydefault}{\mddefault}{\updefault}$N$}}}
\put(  1,-811){\makebox(0,0)[lb]{\smash{\SetFigFont{8}{9.6}{\familydefault}{\mddefault}{\updefault}$\sigma_S$}}}
\put(733,-2999){\makebox(0,0)[b]{\smash{\SetFigFont{8}{9.6}{\familydefault}{\mddefault}{\updefault}0}}}
\put(1498,-2999){\makebox(0,0)[b]{\smash{\SetFigFont{8}{9.6}{\familydefault}{\mddefault}{\updefault}1}}}
\put(2264,-2999){\makebox(0,0)[b]{\smash{\SetFigFont{8}{9.6}{\familydefault}{\mddefault}{\updefault}2}}}
\put(3029,-2999){\makebox(0,0)[b]{\smash{\SetFigFont{8}{9.6}{\familydefault}{\mddefault}{\updefault}3}}}
\put(3795,-2999){\makebox(0,0)[b]{\smash{\SetFigFont{8}{9.6}{\familydefault}{\mddefault}{\updefault}4}}}
\put(4560,-2999){\makebox(0,0)[b]{\smash{\SetFigFont{8}{9.6}{\familydefault}{\mddefault}{\updefault}5}}}
\put(5326,-2999){\makebox(0,0)[b]{\smash{\SetFigFont{8}{9.6}{\familydefault}{\mddefault}{\updefault}6}}}
\put(659,329){\makebox(0,0)[rb]{\smash{\SetFigFont{8}{9.6}{\familydefault}{\mddefault}{\updefault}0.7}}}
\put(6091,-2999){\makebox(0,0)[b]{\smash{\SetFigFont{8}{9.6}{\familydefault}{\mddefault}{\updefault}7}}}
\put(659,-2875){\makebox(0,0)[rb]{\smash{\SetFigFont{8}{9.6}{\familydefault}{\mddefault}{\updefault}0}}}
\put(659,-129){\makebox(0,0)[rb]{\smash{\SetFigFont{8}{9.6}{\familydefault}{\mddefault}{\updefault}0.6}}}
\put(659,-586){\makebox(0,0)[rb]{\smash{\SetFigFont{8}{9.6}{\familydefault}{\mddefault}{\updefault}0.5}}}
\put(659,-1044){\makebox(0,0)[rb]{\smash{\SetFigFont{8}{9.6}{\familydefault}{\mddefault}{\updefault}0.4}}}
\put(659,-1502){\makebox(0,0)[rb]{\smash{\SetFigFont{8}{9.6}{\familydefault}{\mddefault}{\updefault}0.3}}}
\put(659,-1960){\makebox(0,0)[rb]{\smash{\SetFigFont{8}{9.6}{\familydefault}{\mddefault}{\updefault}0.2}}}
\put(659,-2417){\makebox(0,0)[rb]{\smash{\SetFigFont{8}{9.6}{\familydefault}{\mddefault}{\updefault}0.1}}}
\end{picture}
\caption{As Fig.~\protect\ref{fig:bimodal_sigma_0_015}, but with
$q=12$, $x=0.03$.}
\label{fig:bimodal_sigma_0_03}
\efig

Repeating the above calculations for a range of values of $x$, we
arrive at the dashed curve in Fig.~\ref{fig:parent_bimodality}: at a
given $x$, we have three phase coexistence for values of $\ssch$ below
the curve. Conversely, horizontal sections through the graph give us
the extent in $x$ of the three-phase region for given $\ssch$. Since
the three-phase region cannot extend all the way to $x=0$, this shows
that the dashed curve must eventually bend down towards the horizontal
axis as $x$ decreases and meet the axis at some small but nonzero
value of $x$. We have not been able to detect this part of the curve,
however, since the $x$-values concerned are too small for our numerics
to be accurate.

We can now analyse whether there is a correlation between three-phase
coexistence region and bimodality of the parent length
distribution. For values of $x$ and $\ssch$ below the solid curve in
Fig.~\ref{fig:parent_bimodality}, the parent is bimodal; above the
curve we have a unimodal length distribution. The comparison of the
regions under the dashed and solid curves suggests that bimodality may
be required for three-phase coexistence to occur (except possibly in
the region of very small $x$, where our data on three-phase
coexistence are incomplete), but does not necessary entail it.

One might therefore look for additional properties of the length
distribution as predictors of three-phase coexistence, e.g.\ the
overall polydispersity $\sigma$. 
Eq.~\eqref{eq:bidisp_sigma} showed that for a bidisperse system the
overall polydispersity $\sigma$ vanishes at $x=0$ and $x=1$ and has a
maximum for $x=1/(q+1)$; the values of $x$ where three-phase
coexistence occurs in Fig.~\ref{fig:parent_bimodality} are smaller but
at least of the same order as those where $\sigma$ is maximized. Also,
the maximum polydispersity in a bidisperse distribution with a given
length ratio $q$ increases with $q$, in line with the general trend
that separation into three phases requires sufficiently large
$q$. Nevertheless, it is clear that the trend of three-phase
separation occurring preferably for systems with large $\sigma$ cannot
be universally true. In fact, it is easy to see this from
Fig.~\ref{fig:parent_bimodality}. When we increase the width $\ssch$
of the two peaks in a bimodal distribution the overall polydispersity
$\sigma$ also increases; but this actually suppresses three-phase
coexistence rather than enhance it.

Loosely speaking, our results show that, in order for the system to be
able to separate into three phases, we need a sufficiently large
disparity between long and short rods in the system, i.e.\ large $q$;
but at the same time the
concentration of short and long rods must not be too similar. The
second condition would account for the fact that the three-phase
region develops only at small values of $x$, and disappears as $\ssch$
is increased. 
If this interpretation is correct, then we should be able to see
three-phase coexistence even for a {\em unimodal} rod length
distribution, as long as it contains a sufficiently large number of
rods that are much longer than the average. A good candidate for this
is a log-normal distribution, which has been shown already in the
context of polymer solutions to present some interesting features
caused by the presence of very large
particles~\cite{Solc70,Solc75}. We will present our analysis of this
case in a separate publication~\cite{SpeSol_p2_fat}.
\bfig
\begin{picture}(0,0)%
\epsfig{file=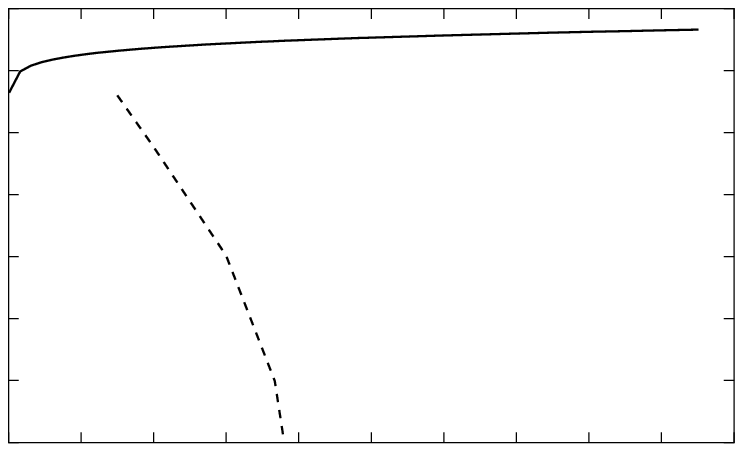}%
\end{picture}%
\setlength{\unitlength}{2565sp}%
\begingroup\makeatletter\ifx\SetFigFont\undefined%
\gdef\SetFigFont#1#2#3#4#5{%
  \reset@font\fontsize{#1}{#2pt}%
  \fontfamily{#3}\fontseries{#4}\fontshape{#5}%
  \selectfont}%
\fi\endgroup%
\begin{picture}(5845,3752)(376,-3286)
\put(376,-811){\makebox(0,0)[lb]{\smash{\SetFigFont{8}{9.6}{\familydefault}{\mddefault}{\updefault}$\sigma$}}}
\put(3301,-3286){\makebox(0,0)[lb]{\smash{\SetFigFont{8}{9.6}{\familydefault}{\mddefault}{\updefault}$x$}}}
\put(659,-1502){\makebox(0,0)[rb]{\smash{\SetFigFont{8}{9.6}{\familydefault}{\mddefault}{\updefault}0.3}}}
\put(659,-1044){\makebox(0,0)[rb]{\smash{\SetFigFont{8}{9.6}{\familydefault}{\mddefault}{\updefault}0.4}}}
\put(659,-586){\makebox(0,0)[rb]{\smash{\SetFigFont{8}{9.6}{\familydefault}{\mddefault}{\updefault}0.5}}}
\put(659,-129){\makebox(0,0)[rb]{\smash{\SetFigFont{8}{9.6}{\familydefault}{\mddefault}{\updefault}0.6}}}
\put(659,329){\makebox(0,0)[rb]{\smash{\SetFigFont{8}{9.6}{\familydefault}{\mddefault}{\updefault}0.7}}}
\put(733,-2999){\makebox(0,0)[b]{\smash{\SetFigFont{8}{9.6}{\familydefault}{\mddefault}{\updefault}0}}}
\put(1269,-2999){\makebox(0,0)[b]{\smash{\SetFigFont{8}{9.6}{\familydefault}{\mddefault}{\updefault}0.01}}}
\put(1805,-2999){\makebox(0,0)[b]{\smash{\SetFigFont{8}{9.6}{\familydefault}{\mddefault}{\updefault}0.02}}}
\put(659,-1960){\makebox(0,0)[rb]{\smash{\SetFigFont{8}{9.6}{\familydefault}{\mddefault}{\updefault}0.2}}}
\put(2340,-2999){\makebox(0,0)[b]{\smash{\SetFigFont{8}{9.6}{\familydefault}{\mddefault}{\updefault}0.03}}}
\put(659,-2875){\makebox(0,0)[rb]{\smash{\SetFigFont{8}{9.6}{\familydefault}{\mddefault}{\updefault}0}}}
\put(2876,-2999){\makebox(0,0)[b]{\smash{\SetFigFont{8}{9.6}{\familydefault}{\mddefault}{\updefault}0.04}}}
\put(3412,-2999){\makebox(0,0)[b]{\smash{\SetFigFont{8}{9.6}{\familydefault}{\mddefault}{\updefault}0.05}}}
\put(3948,-2999){\makebox(0,0)[b]{\smash{\SetFigFont{8}{9.6}{\familydefault}{\mddefault}{\updefault}0.06}}}
\put(4484,-2999){\makebox(0,0)[b]{\smash{\SetFigFont{8}{9.6}{\familydefault}{\mddefault}{\updefault}0.07}}}
\put(5019,-2999){\makebox(0,0)[b]{\smash{\SetFigFont{8}{9.6}{\familydefault}{\mddefault}{\updefault}0.08}}}
\put(5555,-2999){\makebox(0,0)[b]{\smash{\SetFigFont{8}{9.6}{\familydefault}{\mddefault}{\updefault}0.09}}}
\put(6091,-2999){\makebox(0,0)[b]{\smash{\SetFigFont{8}{9.6}{\familydefault}{\mddefault}{\updefault}0.1}}}
\put(659,-2417){\makebox(0,0)[rb]{\smash{\SetFigFont{8}{9.6}{\familydefault}{\mddefault}{\updefault}0.1}}}
\end{picture}
\vspace*{0.1cm}
\caption{Shown is, for a mixture of two Schulz distributions with
length ratio $q=12$, the value of $\ssch$ at which we lose the
bimodality of the parent (solid), together with the maximum value of
$\ssch$ for which we have a three-phase region for the given value of
$x$ (dashed).  The parent is bimodal in the region below the solid
curve.  Note that, from theoretical considerations, the dashed curve
has to bend over as $x$ decreases and eventually meet the
horizontal axis at some small but positive value of $x$.}
\label{fig:parent_bimodality}
\efig

\section{Conclusion}
\label{sec:conclusion}

We have analysed the phase equilibria of the $\Ptwo$ Onsager model of
hard rods with length polydispersity. The model is defined by a
truncation of the angular 
dependence of the 
excluded volume interaction of the Onsager theory after the
second Legendre polynomial $\Ptwo(\cos\theta)$.
Within this model we derived the exact phase equilibrium conditions,
but these are still rather difficult to solve numerically. We therefore
exploited the fact that the model is truncatable -- the excess free energy is 
a function of only two moments ($\rho_1$ and $\rho_2$) of the
density distribution -- and used the moment method, from which
numerical solutions for phase equilibria can be obtained with
well-controlled accuracy; the onset of nematic ordering from the isotropic
side is found exactly by construction of the moment free energy.

For a fully polydisperse case with a unimodal (Schulz) length we found
some of the common features of polydisperse systems. In particular, we
observed a strong fractionation
(Fig.~\ref{fig:fractionation_unimodal}) and a pronounced broadening of
the coexistence region (Fig.~\ref{fig:phase_diag_unimodal}) with
increasing polydispersity; the latter is defined as the normalized
standard deviation of the length distribution.  Fractionation effects
were strong enough to lead to a crossing of the cloud point curves
with the corresponding shadow curves in the number density
representation (Fig.~\ref{fig:phase_diag_unimodal}), but not in the
volume fraction representation
(Fig.~\ref{fig:vol_frac_unimodal}). This implies that while the
nematic phase always has a larger rod volume fraction than the
isotropic, it can actually have a smaller rod number density.

For the unimodal length distribution we did not observe any
three-phase I-N-N or two-phase N-N coexistence, or re-entrant behaviour
of the nematic phase, all of which had been found in the Onsager model
with bi- and tridisperse length distributions. In order to understand
whether this was due to the introduction of a continuous and unimodal
length distribution, or to the truncation that defined the $\Ptwo$
Onsager model, we next studied the bidisperse case within the $\Ptwo$
model.  As the length ratio $q$ increased, we indeed found
(Fig.~\ref{fig:bidisp_q_3_5_both} and~\ref{fig:bidisp_q_7_both}) the
development of a re-entrance of the nematic phase, and eventually also
I-N-N and N-N coexistence (Fig.~\ref{fig:bidisperse_phase_diag}) for
larger $q$. Thus all the qualitative features of the bidisperse phase
diagrams of the full Onsager theory are recovered. The only exception
is the presence in the $\Ptwo$ model of a critical point that closes
the N-N coexistence region at high density, while for the full Onsager
theory it is known~\cite{VanMul96} that for large $q$ such a critical
point does not exist. We explained this difference in terms of the
different behaviour of the excluded volume interaction in the limit of
strongly ordered rods, and argued that it would persist for any
truncation of the full Onsager theory involving only a finite number
of Legendre polynomials.

Finally, in order to interpolate between the different behaviours in
the unimodal and bidisperse cases, we introduced polydispersity into
the latter system, by considering mixtures of two Schulz distributions
of width $\ssch$ peaked at different rod lengths; the bidisperse
system is recovered for $\ssch\to 0$.  For small $\ssch$, i.e., small
``broadening'' of the bidisperse ``peaks'', the phase diagrams
remained essentially as in the bidisperse limit, while larger values
of $\ssch$ lead to a disappearance of the I-N-N and N-N coexistence
regions (Fig.~\ref{fig:phased_q_12_sigma_0.6}).  It is natural to
suppose that this change is related to the fact that the length
distribution changes from bimodal to unimodal as $\ssch$ is
increased. However, a closer investigation of this point showed that
bimodality, although correlated with the occurrence of I-N-N and N-N
coexistence, is not sufficient to guarantee that such coexistence will
occur. Indeed, this follows already from the fact that in the
bidisperse system (which is always bimodal) a minimum value of the
length ratio $q$ is needed for three-phase coexistence.  Our results
suggest that, also in the polydisperse case, a large ratio between the
typical (e.g., mean) values of the two Schulz length distributions is
required.  However, the result that three-phase coexistence generally
only occurs for small values values of the number fraction $x$ of long
rods suggests that in addition the length distribution must be rather
asymmetric, with fewer long rods than short ones. When the width
$\ssch$ of the Schulz distributions is increased too much, this (here
only vaguely defined) ``asymmetry'' is lost, and the three-phase
region disappears. In order to assess whether this intuition is
correct, we are currently investigating the phase behaviour for
log-normal length distributions~\cite{SpeSol_p2_fat}; while unimodal,
these contain rods of very dissimilar lengths and are also strongly
asymmetric. The intuition developed above would lead one to
expect three-phase coexistence.

We did not investigate in detail the conditions for occurrence of the
re-entrance in the nematic phase boundary. This phenomenon seemed
rather robust compared to I-N-N and N-N coexistence, although in the
unimodal (Schulz distribution) case it was again absent. In the
bidisperse case the re-entrance appeared at lower values of $q$ than
the three-phase coexistence, and also did not require the asymmetry
$x$ between long and short rod number densities to be quite so
extreme, extending e.g.\ almost up to $x=0.3$ for $q=12$
(Fig.~\ref{fig:bidisperse_phase_diag}). In the case of a mixture of
Schulz distributions case it also survived to larger values of the width
$\ssch$ of the two peaks of the distribution. An intriguing observation is
that, for the bidisperse systems, the re-entrance begins (starting from
small $x$) very close to the value of $x=1/(q+1)$, where the
polydispersity $\sigma$ of the system is maximal; see Eq.~\eqref{eq:bidisp_sigma}.
This is also the value of $x$ where the volume fractions occupied by the
short and long rods become equal; see Eq.~\eqref{eq:bidisp_volfract}. We have not been able 
to find a simple physical explanation for this observation.

In summary, our study of the $\Ptwo$ Onsager model shows that the
details of the rod length distribution can have profound effects on
the phase behaviour. In particular, while bidisperse systems can show
a re-entrant nematic phase boundary as well as I-N-N and N-N
coexistence, these features are absent from the phase diagram for a
unimodal length distribution; they also disappear as one interpolates
between the bidisperse and unimodal extremes. Put differently, the
more ``exotic'' aspects of the phase behaviour of mixtures of hard
rods are not expected to be observed in systems with only a moderate
spread of rod lengths around a mean value, but require sufficiently
wide (and, as we argued above, asymmetric) rod length distributions.
Our comparison of the predictions of the $\Ptwo$ model and the full
Onsager theory for the bidisperse case suggests that these conclusions
are not artefacts of the truncation used to the construct the $\Ptwo$
model. Rather, we expect that they would qualitatively also be found
in an analysis of the effects of length polydispersity in the
framework of the full Onsager model; such an analysis remains a
challenging problem for future work.

We comment finally on the presence of the N-N critical point within
the bidisperse $\Ptwo$ Onsager model. As discussed above, the theoretical
work of van Roij and Mulder~\cite{VanMul96} shows that for
sufficiently large ratios $q$ of rod lengths such a critical point
will not occur in the full Onsager theory; earlier reports to the
contrary~\cite{BirKolPry88} appear to be artefacts of numerical
approximations. However, the results of Ref.~\cite{VanMul96} show only
that the N-N coexistence region is open, i.e., extends to arbitrarily
large densities, for $q$ above $\approx 3.17$. The possibility remains
that for a small range of $q$-values below 3.17 an N-N coexistence
region would already exist, at moderate densities and closed off by a
critical point; the topology of the phase diagram for such $q$ would
be akin to the $\Ptwo$ Onsager phase diagrams that contain N-N and
I-N-N coexistence regions. As $q$ increases, the critical point would
then have to move to larger and larger densities so that the N-N
region eventually becomes open at $q\approx 3.17$. It would be worth
revisiting the bidisperse Onsager model to confirm this scenario.

{\em Acknowledgement:} PS acknowledges financial support through EPSRC
grant GR/R52121/01.

\appendix

\section{The strong nematic ordering approximation}
\label{sec:SNO_approx}

In this appendix we describe an approximation that is useful for
analysing the phase diagram of the $\Ptwo$ Onsager model at high
densities. This strong nematic ordering (SNO) approximation becomes
exact in the limit $\rho_0\to\infty$ and allows one to show formally
that there can be no N-N coexistence in this limit. For finite
$\rho_0$ it is approximate but nevertheless useful for understanding
some limits of the phase behaviour, e.g.\ for large values of the rod
length ratio $q$ in bidisperse and bimodal length distributions. As
discussed in Sec.~\ref{sec:bidisperse_res}, it is in the regime of
strong nematic ordering (small angles $\theta$) that the $\Ptwo$
Onsager model and the full Onsager theory differ most strongly, so the
analysis here focusses on the properties that are
particular to the $\Ptwo$ Onsager model.

The SNO approximation is based on the assumption that all rod angles
$\theta$ in any nematic phases are reasonably small, so that we can
approximate $\cos^2\theta\simeq 1-\theta^2$ and hence
\beq\label{eq:P2_approx}
\Ptwo(\cos\theta)\simeq 1-\dfrac{3}{2}\theta^2
\eeq
From Eq.~\eqref{eq:P_l}, the orientational distributions are then
$P_l(\theta) \sim \exp[-\frac{3}{2}lc_2\rho_2\theta^2]$ and so the SNO
assumption of small angles $\theta$ corresponds to $\rho_2$ being
large enough to have $l\rho_2\gg 1$ for all lengths $l$ that contribute
significantly. Using the approximation~\eqref{eq:P2_approx}, we can
get explicit expressions for the length dependent chemical
potentials. The relevant angular integral can be evaluated as
\bea\label{eq:angint_approx}
\nonumber \angint e^{lc_2\rho_2\Ptwo}
&\simeq&\int_0^{\pi/2}d\theta\,\sin\theta\, 
e^{lc_2\rho_2-(3/2)lc_2\rho_2\theta^2}\\
&\simeq&\dfrac{\exp\left(lc_2\rho_2\right)}{3lc_2\rho_2}
%\left(1-\exp\left(-\frac{3}{2}lc_2\rho_2\dfrac{\pi^2}{4}\right)\right)
\eea
where the symmetry of the integrand under $\theta\to\pi-\theta$ has
been exploited and corrections that are exponentially small in
$l\rho_2$ have been neglected. The chemical potentials~\eqref{eq:mu}
thus become
\beq
\label{eq:approx_chem_pot}
\label{eq:muN}
\mu\N(l)\simeq\ln\rho(l)+l(c_1\rho_1-c_2\rho_2)+\ln(3lc_2\rho_2)
%-\ln\left(1-\exp\left(-\frac{3}{2}lc_2\rho_2\dfrac{\pi^2}{4}\right)\right)
\end{equation}
for a nematic phase, while for an isotropic phase ($\rho_2=0$) the
angular integral~\eqref{eq:angint_approx} equals unity and the
chemical potentials are
\beq
\mu\I(l)=\ln\rho(l)+lc_1\rho_1\label{eq:muI}
%\mu\N(l)&=&\ln\rho\N(l)+l(c_1\rho\N_1-c_2\rho\N_2)+\ln(3lc_2\rho\N_2)
\eeq
Using the SNO assumption~\eqref{eq:P2_approx} we can also evaluate the
integrals appearing in the self-consistency equation~\eqref{eq:rho2}
for $\rho_2$; the latter can then be solved explicitly to give,
after some algebra, 
\beq\label{eq:rho2_approx}
\rho_2\simeq\dfrac{\rho_1}{2}
\left(1+\sqrt{1-\dfrac{4\rho_0}{c_2\rho_1^2}}\right)
\end{equation}
In the limit of large density $\rho_0$, since $\rho_1=\langle l\rangle
\rho_0 \sim \rho_0$, the square root approaches unity and thus
$\rho_2 \simeq \rho_1$. This is as expected, since the weight
functions~(\ref{eq:w1},\ref{eq:w2}) for $\rho_1$ and $\rho_2$ only
differ by a factor of $\Ptwo(\cos\theta)$ which becomes equal to one
for maximal nematic order, i.e.\ for $\theta\to 0$. The equality
$\rho_2=\rho_1$ implies that $\rho_2$ becomes large as the density
increases; for $\rho_0\to\infty$ it diverges and the SNO approximation
becomes exact. One can now easily show that in this large density
limit there can be no coexistence between two nematic phases. The
osmotic pressure~\eqref{eq:mom_Pi} becomes
$\Pi=\rho_0+\frac{1}{2}(c_1-c_2)\rho_1^2\simeq
\frac{1}{2}(c_1-c_2)\rho_1^2$. Coexisting nematics would therefore
need to have the same $\rho_1$, hence also the same $\rho_2$. But then
the excess parts of the chemical potentials $\mu(l)$,
Eq.~\eqref{eq:mu}, would also be the same in the two phases. Chemical
potential equality then also forces equality of the density
distributions $\rho(l)$, so that the supposed coexisting nematics turn out to
be the same phase.

Using the SNO approximation~(\ref{eq:muN},\ref{eq:rho2_approx}) we can
furthermore get approximate results for the nematic cloud curves, i.e.\
the phase boundaries at high densities between the single-phase
nematic region and any I-N or N-N coexistence regions. The single
nematic phase must have the density distribution of the parent,
$\rho(l)=\rho_0\normparent(l)$, with $\rho_1=\rho_0$ since the parent
distribution has unit average rod length; its value of $\rho_2$ is
given by~\eqref{eq:rho2_approx}. We need to find the value of the
density $\rho_0$ where this phase first begins to coexist with a new
isotropic or nematic phase as $\rho_0$ is lowered. For I-N
coexistence, equality of the chemical
potentials~(\ref{eq:muN},\ref{eq:muI}) implies that the isotropic
phase has density distribution
\beq\label{eq:I_coexist_approx}
\rho\I(l)=3lc_2\rho_2\rho_0e^{l\beta}P^{(0)}(l)
\eeq
with
\beq\label{eq:beta_I-N}
\beta=c_1(\rho_0-\rho_1\I)-c_2\rho_2
\eeq
For given $\beta$ and $\rho_0$, $\rho_1\I$ is given by $\rho_1\I =
\int dl\, l\rho\I(l)$ and the onset of I-N phase coexistence can
thus be determined simply by solving numerically for the values of
$\beta$ and $\rho_0$ that fulfill Eq.~\eqref{eq:beta_I-N} and give
equal osmotic pressure~\eqref{eq:mom_Pi} in the two phases.

For N-N coexistence one similarly finds that the new nematic phase has
density distribution
\beq\label{eq:N_coexist_approx}
\rho\N(l)=\dfrac{\rho_2}{\rho_2\N}\rho_0 e^{l\beta}\normparent(l)
\end{equation}
with
\beq\label{eq:beta_N-N}
\beta=c_1(\rho_0-\rho_1\N)-c_2(\rho_2-\rho_2\N)
\eeq
and one can again solve numerically, e.g.\ for $\beta$, $\rho_0$,
$\rho\N_0$ and $\rho\N_1$. (The four conditions are then
Eq.~\eqref{eq:beta_N-N}, $\rho_0\N=\int dl\ \rho\N(l)$, $\rho_1\N=\int
dl\ l \rho\N(l)$, and the osmotic pressure equality; $\rho\N_2$ is
given by Eq.~\eqref{eq:rho2_approx} in terms of $\rho_0\N$ and
$\rho\N_1$.)

We can obtain in this way approximate nematic cloud curves for
bidisperse and bimodal parent length distributions; these reproduce
qualitatively the features of our numerically exact phase diagrams,
including the re-entrance of the I-N phase boundary. Whether I-N or
N-N phase separation actually occurs as density is lowered depends on
which of the calculated cloud curves is reached first; due to its
overestimate of the nematic ordering, it turns out the SNO predicts
stable N-N demixing only above a larger threshold value of the length
ratio $q$ than in the numerically exact phase diagrams. The overall
trends are, however, the same as in the numerically exact
phase diagrams of Sec.~\ref{sec:bidisperse_res} and
\ref{sec:bimodal_res}. Increasing $q$ increases the size of the
re-entrant region of the I-N phase boundary; the N-N phase boundary
also moves to larger densities and becomes stable over a wider
range. Increasing the peak width $\ssch$, on the other hand, produces
the opposite effects. Interestingly, the SNO predicts that the I-N
phase boundary remains re-entrant even for the largest values of the
peak width, $\ssch=1$. We were not able to check this prediction by
calculating exact phase diagrams for $\ssch=1$, but it does suggest that the
re-entrance is mainly due to the overall polydispersity of the parent,
rather than the presence of two well separated peaks in the length
distribution.

On the N-N phase boundaries that we calculate within the SNO
approximation, a critical point occurs, signalled by the vanishing of
$\beta$, Eq.~\eqref{eq:beta_N-N}, and the coincidence of all moments
$\rho_0$, $\rho_1$, $\rho_2$ in the two phases. This critical point is
not always observable, however, since it can lie in the metastable
region inside the I-N phase boundary. This can happen even if other
parts of the N-N phase boundary {\em are} stable. We find that this
situation occurs for a narrow range of $q$, for genuinely polydisperse
(bimodal rather than bidisperse) distributions. This prediction of the
SNO approximation may appear peculiar but must be expected to be a
general feature of systems containing more than two types of rods. Our
($\rho_0,x$) phase diagrams are then planar cuts through a
higher-dimensional phase diagram; tielines beginning in the cut plane
will connect to coexisting phases off the plane. For an appropriately
chosen plane, all tielines originating from a stable N-N phase
boundary in the plane may then point ``to one side'' of the plane,
with none of them having zero length and thus giving a critical point.

Finally, we use the SNO approximation to analyse the phase diagrams
for bidisperse and bimodal length distributions in the limit of large
rod length ratio $q$; this limit would be difficult to access using
direct numerical calculations of the phase diagrams that do not employ
the SNO assumption. We also expect the SNO approximation to become
more accurate for larger $q$. This is suggested by the fact that
an increase in $q$ generally causes the nematic cloud point 
to shift to higher density; at such higher densities the nematic will
be more strongly ordered, and thus the SNO approximation better justified.
While such an argument applies directly to the case of the
boundary between the nematic region and the I-N region, the same
cannot be said for the N-N phase boundary. There the hypothesis of
strong ordering is applied to both nematic phases in the SNO approach,
and there is no trivial reason why a nematic shadow phase coexisting with a
denser nematic parent is necessarily strongly ordered. Numerically, however, we
have found this trend to be confirmed.

It turns out that in the limit $q\to\infty$ the equations that we need
to solve to obtain the nematic cloud curves, i.e.\ the onset of phase
separation into I-N or N-N when decreasing the density, become
dependent only on the scaling variable $\xi=xq$ rather than
separately on $q$ and the fraction of long rods $x$. This
simplification occurs because terms that are exponentially small in
$q$ can be neglected for $q\to\infty$. Intuitively, $\xi$ is for large
$q$ simply the ratio of the volume fractions of long and short rods.
This follows from the fact that the ratio of number densities is
$x/(1-x)$, so that the volume fractions have the ratio
\beq\label{eq:bidisp_volfract}
\frac{l_2x}{l_1(1-x)}=\frac{qx}{1-x}
\eeq
which tends to $\xi$ for $x=\xi/q\to0$.
The numerical solution of the resulting simplified system of equations
shows that there is always a stable region of N-N demixing for
$q\to\infty$; interestingly, this is true even for bimodal
distributions with the largest peak width $\ssch=1$ that we
consider. As explained in Sec.~\ref{sec:bimodal_res}, this is presumably due to
the fact that in the limit $q\to\infty$ (with $x=\xi/q$ and $\xi$ held
constant) the parent always has two separate peaks, whatever the value
of $\ssch$.
%
%\vspace*{-0.3cm}
\bfig
\begin{picture}(0,0)%
\epsfig{file=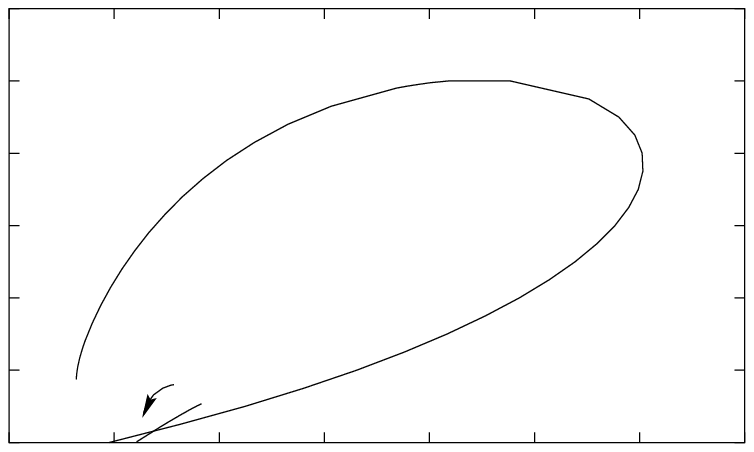}%
\end{picture}%
\setlength{\unitlength}{2565sp}%
\begingroup\makeatletter\ifx\SetFigFont\undefined%
\gdef\SetFigFont#1#2#3#4#5{%
  \reset@font\fontsize{#1}{#2pt}%
  \fontfamily{#3}\fontseries{#4}\fontshape{#5}%
  \selectfont}%
\fi\endgroup%
\begin{picture}(6190,3752)(1,-3286)
\put(1726,-2311){\makebox(0,0)[lb]{\smash{\SetFigFont{8}{9.6}{\familydefault}{\mddefault}{\updefault}$N_1+N_2$}}}
\put(5251,-1936){\makebox(0,0)[lb]{\smash{\SetFigFont{8}{9.6}{\familydefault}{\mddefault}{\updefault}$N$}}}
\put(2776,-1336){\makebox(0,0)[lb]{\smash{\SetFigFont{8}{9.6}{\familydefault}{\mddefault}{\updefault}$I+N$}}}
\put(  1,-511){\makebox(0,0)[lb]{\smash{\SetFigFont{8}{9.6}{\familydefault}{\mddefault}{\updefault}$\xi$}}}
\put(901,-1336){\makebox(0,0)[lb]{\smash{\SetFigFont{8}{9.6}{\familydefault}{\mddefault}{\updefault}$I$}}}
\put(3301,-3286){\makebox(0,0)[lb]{\smash{\SetFigFont{8}{9.6}{\familydefault}{\mddefault}{\updefault}$\rho_0$}}}
\put(659,-2999){\makebox(0,0)[b]{\smash{\SetFigFont{8}{9.6}{\familydefault}{\mddefault}{\updefault}0}}}
\put(1435,-2999){\makebox(0,0)[b]{\smash{\SetFigFont{8}{9.6}{\familydefault}{\mddefault}{\updefault}5}}}
\put(2211,-2999){\makebox(0,0)[b]{\smash{\SetFigFont{8}{9.6}{\familydefault}{\mddefault}{\updefault}10}}}
\put(2987,-2999){\makebox(0,0)[b]{\smash{\SetFigFont{8}{9.6}{\familydefault}{\mddefault}{\updefault}15}}}
\put(3763,-2999){\makebox(0,0)[b]{\smash{\SetFigFont{8}{9.6}{\familydefault}{\mddefault}{\updefault}20}}}
\put(4539,-2999){\makebox(0,0)[b]{\smash{\SetFigFont{8}{9.6}{\familydefault}{\mddefault}{\updefault}25}}}
\put(585,329){\makebox(0,0)[rb]{\smash{\SetFigFont{8}{9.6}{\familydefault}{\mddefault}{\updefault}12}}}
\put(5315,-2999){\makebox(0,0)[b]{\smash{\SetFigFont{8}{9.6}{\familydefault}{\mddefault}{\updefault}30}}}
\put(585,-2875){\makebox(0,0)[rb]{\smash{\SetFigFont{8}{9.6}{\familydefault}{\mddefault}{\updefault}0}}}
\put(6091,-2999){\makebox(0,0)[b]{\smash{\SetFigFont{8}{9.6}{\familydefault}{\mddefault}{\updefault}35}}}
\put(585,-205){\makebox(0,0)[rb]{\smash{\SetFigFont{8}{9.6}{\familydefault}{\mddefault}{\updefault}10}}}
\put(585,-739){\makebox(0,0)[rb]{\smash{\SetFigFont{8}{9.6}{\familydefault}{\mddefault}{\updefault}8}}}
\put(585,-1273){\makebox(0,0)[rb]{\smash{\SetFigFont{8}{9.6}{\familydefault}{\mddefault}{\updefault}6}}}
\put(585,-1807){\makebox(0,0)[rb]{\smash{\SetFigFont{8}{9.6}{\familydefault}{\mddefault}{\updefault}4}}}
\put(585,-2341){\makebox(0,0)[rb]{\smash{\SetFigFont{8}{9.6}{\familydefault}{\mddefault}{\updefault}2}}}
\end{picture}
\caption{Phase diagram as derived
from the SNO approximation, for a bidisperse system with $q\to\infty$;
the scaling variable on the vertical axis is $\xi=xq$. The straight
line near the bottom left corner of the plot indicates the onset of
N-N coexistence, which is stable -- since it occurs at higher density
than I-N phase separation -- at sufficiently small $\xi$. The same is
still true if we move to a bimodal distribution, even for the maximal
peak widths $\ssch=1$ that we consider.}
\label{fig:bid_asy}
\efig
The N-N phase boundary moves towards larger densities as $q$
increases, and in fact becomes remarkably simple as $q\to\infty$: the
N-N critical point moves towards $\xi=0$, and the part of the N-N
phase boundary at densities below the critical point collapes onto the
horizontal axis. A plot of the N-N cloud curve for different $q$ with
the scaling quantity $\xi=xq$ on the vertical axis shows this
evolution explicitly (Fig.~\ref{fig:NN_asy}). Note that in the same
representation the I-N phase boundary would change rather little for the
same range of values of $q$, showing that the relative stability of
the N-N phase separation (compared to I-N) is enhanced for large
$q$. This is consistent with the trend observed in the numerically
exact phase diagrams on Sec.~\ref{sec:bidisperse_res}.
%
%\vspace*{-0.3cm}
\bfig
\begin{picture}(0,0)%
\epsfig{file=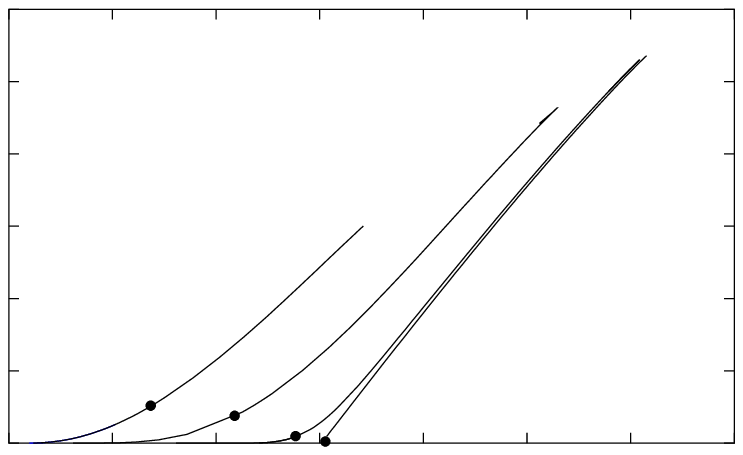}%
\end{picture}%
\setlength{\unitlength}{2565sp}%
\begingroup\makeatletter\ifx\SetFigFont\undefined%
\gdef\SetFigFont#1#2#3#4#5{%
  \reset@font\fontsize{#1}{#2pt}%
  \fontfamily{#3}\fontseries{#4}\fontshape{#5}%
  \selectfont}%
\fi\endgroup%
\begin{picture}(6040,3881)(151,-3415)
\put(4351,-1411){\makebox(0,0)[lb]{\smash{\SetFigFont{8}{9.6}{\familydefault}{\mddefault}{\updefault}$q=\infty$}}}
\put(3451,-3361){\makebox(0,0)[lb]{\smash{\SetFigFont{8}{9.6}{\familydefault}{\mddefault}{\updefault}$\parent_0$ }}}
\put(3076,-1186){\makebox(0,0)[lb]{\smash{\SetFigFont{8}{9.6}{\familydefault}{\mddefault}{\updefault}$q=20$}}}
\put(151,-586){\makebox(0,0)[lb]{\smash{\SetFigFont{8}{9.6}{\familydefault}{\mddefault}{\updefault}$\xi$}}}
\put(5101, 89){\makebox(0,0)[lb]{\smash{\SetFigFont{8}{9.6}{\familydefault}{\mddefault}{\updefault}$q=1000$}}}
\put(4426,-286){\makebox(0,0)[lb]{\smash{\SetFigFont{8}{9.6}{\familydefault}{\mddefault}{\updefault}$q=70$}}}
\put(733,-2999){\makebox(0,0)[b]{\smash{\SetFigFont{8}{9.6}{\familydefault}{\mddefault}{\updefault}3}}}
\put(1498,-2999){\makebox(0,0)[b]{\smash{\SetFigFont{8}{9.6}{\familydefault}{\mddefault}{\updefault}4}}}
\put(2264,-2999){\makebox(0,0)[b]{\smash{\SetFigFont{8}{9.6}{\familydefault}{\mddefault}{\updefault}5}}}
\put(3029,-2999){\makebox(0,0)[b]{\smash{\SetFigFont{8}{9.6}{\familydefault}{\mddefault}{\updefault}6}}}
\put(3795,-2999){\makebox(0,0)[b]{\smash{\SetFigFont{8}{9.6}{\familydefault}{\mddefault}{\updefault}7}}}
\put(4560,-2999){\makebox(0,0)[b]{\smash{\SetFigFont{8}{9.6}{\familydefault}{\mddefault}{\updefault}8}}}
\put(659,329){\makebox(0,0)[rb]{\smash{\SetFigFont{8}{9.6}{\familydefault}{\mddefault}{\updefault}1.2}}}
\put(5326,-2999){\makebox(0,0)[b]{\smash{\SetFigFont{8}{9.6}{\familydefault}{\mddefault}{\updefault}9}}}
\put(659,-2875){\makebox(0,0)[rb]{\smash{\SetFigFont{8}{9.6}{\familydefault}{\mddefault}{\updefault}0}}}
\put(6091,-2999){\makebox(0,0)[b]{\smash{\SetFigFont{8}{9.6}{\familydefault}{\mddefault}{\updefault}10}}}
\put(659,-205){\makebox(0,0)[rb]{\smash{\SetFigFont{8}{9.6}{\familydefault}{\mddefault}{\updefault}1}}}
\put(659,-739){\makebox(0,0)[rb]{\smash{\SetFigFont{8}{9.6}{\familydefault}{\mddefault}{\updefault}0.8}}}
\put(659,-1273){\makebox(0,0)[rb]{\smash{\SetFigFont{8}{9.6}{\familydefault}{\mddefault}{\updefault}0.6}}}
\put(659,-1807){\makebox(0,0)[rb]{\smash{\SetFigFont{8}{9.6}{\familydefault}{\mddefault}{\updefault}0.4}}}
\put(659,-2341){\makebox(0,0)[rb]{\smash{\SetFigFont{8}{9.6}{\familydefault}{\mddefault}{\updefault}0.2}}}
\end{picture}
\caption{The N-N phase boundary for bidisperse rod length
distributions with a range of rod length ratios $q$, as derived from
the SNO approximation, with the scaling variable $\xi=xq$ on the
vertical axis. The N-N critical point moves towards $\xi=0$ for
$q\to\infty$, while the N-N boundary as a whole moves towards higher
densities in the same limit.}
\label{fig:NN_asy} 
\efig
%
%\vspace*{-1.2cm}
\bibliographystyle{unsrt}
\bibliography{references}

\begin{thebibliography}{10}
\bibitem{bidisperse_note}
In the bidisperse case we do not need to add any extra moments in our numerical
  calculation, since conservation of the moments $\rho_0$ and $\rho_1$ which we
  include guarantees that the lever rule is fulfilled.
\end{thebibliography}

\end{document}